\pdfoutput=1 

\documentclass[cits]{PoS}
\usepackage{amsmath, amsthm, amssymb}
\usepackage{epsfig}
\usepackage{graphicx}
\usepackage{psfrag}

\def\CO{{\mathcal{O}}}
\def\bar{\overline}

\title{The CKM matrix and flavor physics from lattice QCD}

\ShortTitle{The CKM matrix and flavor physics from lattice QCD}

\author{\speaker{Ruth S. Van de Water}\\
        Physics Department, Brookhaven National Laboratory, Upton, NY 11973\\
        E-mail: \email{ruthv@bnl.gov}}

\abstract{I discuss the role of lattice QCD in testing the Standard Model and searching for physics beyond the Standard Model in the quark flavor sector.  I first review the Standard Model CKM framework.  I then present the current status of the CKM matrix, focusing on determinations of CKM matrix elements and constraints on the CKM unitarity triangle that rely on lattice QCD calculations of weak matrix elements.  I also show the potential impact of improved lattice QCD calculations on the global CKM unitarity triangle fit.  I then describe several hints of new physics in the quark flavor sector that rely on lattice QCD calculations of weak matrix elements, such as evidence of a $\sim 2$--$3 \sigma$ tension in the CKM unitarity triangle and the ``$f_{D_s}$ puzzle''.  I finish with a discussion of lattice QCD calculations of rare $B$- and $K$-decays needed to probe physics beyond the Standard Model at future experiments.}

\FullConference{The XXVII International Symposium on Lattice Field Theory - LAT2009\\
		 July 26-31 2009\\
		 Peking University, Beijing, China}

\begin{document}

\section{Introduction}

The Standard Model of particle physics has proven to be a remarkably successful theory; it describes the observed elementary particles and their interactions and explains the outcomes of high-energy collider experiments.  Nevertheless, we know that the Standard Model is not a ``theory of everything'' because it does not account for neutrino masses, the amount of dark matter and dark energy in the universe, or the abundance of matter over antimatter.  Nor does it explain the origin of particle masses and mixings.  Specifically, in the quark flavor sector, we do not know why there are three generations of quarks, or what generates the hierarchy of quark masses and CKM mixing matrix elements.  Thus we expect that the Standard Model is only a low-energy effective theory, and that we will discover new physics at higher energies that accounts for phenomena such as dark matter and explains the values of Standard Model parameters such as the quark masses and mixings.

\begin{figure}
\begin{center}
\includegraphics[width= 0.7\linewidth]{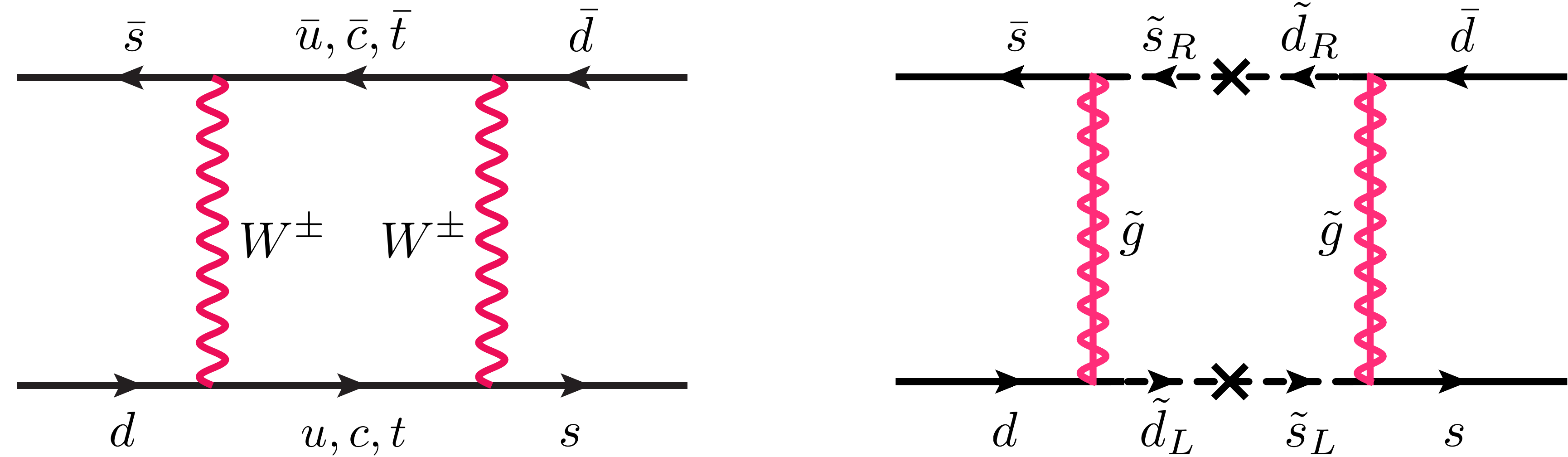}
\caption{Standard Model contribution to $K^0$-$\bar{K^0}$ mixing (left) and a possible supersymmetric contribution (right).}
\label{fig:KKbar}
\end{center}
\end{figure}

Most extensions of the Standard Model contain new $CP$-violating phases and new quark flavor-changing interactions.  We therefore expect new physics effects in the quark flavor sector.  The flavor sector is sensitive to physics at very high scales because new particles will typically appear in loop-level processes.  For example, Fig.~\ref{fig:KKbar} shows a sample supersymmetric contribution to neutral kaon mixing.
Thus we may see evidence for new physics in the flavor sector before we produce non-Standard Model particles directly at the LHC.

Flavor factories such as those at BEPC, Cornell, Frascati, KEK, and SLAC have been pouring out data to measure hadron masses and mixings, branching fractions, and other observables.  These results can be used to obtain CKM matrix elements, in some cases with percent-level accuracy.   We can also compare the measured results with Standard Model predictions; significant inconsistencies would indicate the presence of new physics.  Lattice QCD calculations are needed, however, to interpret many of the results of flavor physics experiments and to obtain the values of the CKM matrix elements.  This is because in order to accurately describe weak interactions involving quarks, one must include effects of confining quarks into hadrons.  Typically the nonperturbative QCD effects are absorbed into quantities such as decay constants, form factors, and bag-parameters that are computed in lattice QCD.  

In this review I present the status of lattice QCD calculations needed to determine CKM matrix elements and constrain the apex of the CKM unitarity triangle.  Towards this aim, I highlight quantities either where significant computational progress has been made in the past few years, or where there is an apparent tension with determinations from independent (non-lattice) methods.  Although I intentionally omit most details about the lattice calculations, I provide references throughout this work for the interested reader.  I also discuss some additional lattice QCD calculations that are of interest to flavor physics phenomenology, but for the most part have not been attempted.  I focus on calculations of interest to rare $B$- and $K$-decays.  I emphasize, however, that this list is by no means exhaustive, and that more lattice QCD calculations of weak matrix elements will become possible as computational resources increase and lattice QCD methods improve.

This paper is organized as follows.  First, in Sec.~\ref{sec:CKM}, I briefly review the Standard Model CKM framework.  Next, in Sec.~\ref{sec:CKM_status}, I summarize the current status of the CKM matrix, focusing on determinations of CKM matrix elements and constraints on the CKM unitarity triangle that rely on inputs from lattice QCD.  In the first subsection, Sec.~\ref{sec:1st_row}, I present determinations of the elements of the first row of the CKM matrix;  the values are currently consistent with the Standard Model expectation of unitarity.  In the second subsection, Sec.~\ref{sec:CKM_UT}, I present current averages for lattice QCD inputs needed to constrain the CKM unitarity triangle.  I also show the potential impact of improved lattice QCD calculations of these quantities on the global CKM unitarity triangle fit.  Next, in Sec.~\ref{sec:NP_hints}, I discuss several hints of new physics in the flavor sector that rely on lattice QCD calculations of weak matrix elements.  For example, given current theoretical and experimental inputs to the CKM unitarity triangle analysis, there is a $\sim 2$--$3\sigma$ tension between observations and Standard Model expectations.  In Sec.~\ref{sec:LQCD_rare}, I discuss prospects for lattice QCD calculations to search for evidence of new physics in rare $B$- and $K$-decays.  Finally, I conclude by discussing future prospects for lattice QCD to aid in new physics searches in the quark flavor sector in Sec.~\ref{sec:Conc}.

\section{The Standard Model CKM framework}
\label{sec:CKM}

In this section I briefly review the Standard Model CKM framework.  For a more thorough discussion, see, for example, Ref.~\cite{Antonelli:2009ws}.

\bigskip

The $3 \times 3$ Cabibbo-Kobayashi-Maskawa (CKM) matrix parameterizes the mixing between quark flavors under weak interactions \cite{Cabibbo:1963yz,Kobayashi:1973fv}:
\begin{equation}
V_{CKM} = \left( \begin{array}{ccc}
     		 V_{ud}  & V_{us} & V_{ub} \\*
     		 V_{cd} & V_{cs} & V_{cb}   \\*
		 V_{td}  & V_{ts}  & V_{tb}  
\end{array} \right)
	= \left( \begin{array}{ccc}
     		 0.9742  & 0.2257 & 3.59 \times10^{-3} \\*
     		 0.2256 & 0.9733 & 41.5 \times 10^{-3}   \\*
		 8.74 \times 10^{-3}  & 40.7 \times 10^{-3}  & 0.9991 
\end{array} \right) ,
\end{equation}
where the numerical values are from Ref.~\cite{Amsler:2008zzb}.  The matrix elements are empirically largest along the diagonal, so mixing is most
probable within the same generation.  Because the CKM matrix is unitary in the Standard Model, one can express elements of the CKM matrix as an expansion in powers of the small parameter $\lambda = |V_{us}| \sim 0.22$.  The resulting Wolfenstein parameterization~\cite{Wolfenstein}, 
\begin{equation}
	V_{CKM} = \left( \begin{array}{ccc}
     		 V_{ud}  & V_{us} & V_{ub} \\*
     		 V_{cd} & V_{cs} & V_{cb}   \\*
		 V_{td}  & V_{ts}  & V_{tb}  
\end{array} \right) 
	= \left( \begin{array}{ccc}
     		 1 -\frac{1}{2}\lambda^2  & \lambda & A \lambda^3(\rho - i\eta) \\*
     		 -\lambda & 1 -\frac{1}{2}\lambda^2 & A \lambda^2   \\*
		 A \lambda^3(1 - \rho - i\eta)  & -A \lambda^2  & 1  
\end{array} \right) + {\cal{O}}(\lambda^4) ,
\end{equation}
where
\begin{equation}
\lambda \equiv |V_{us}|, \qquad A \equiv \frac{|V_{cb}|}{\lambda^2}, \qquad \rho \equiv \frac{\textrm{Re} (V_{ub})}{A \lambda^3}, \qquad \eta \equiv - \frac{\textrm{Im} (V_{ub})}{A \lambda^3} ,
\end{equation}
makes the hierarchy of sizes among the matrix elements explicit in terms of powers $\lambda$.

Because the CKM matrix elements are fundamental parameters of the Standard Model, it is important to know their values as precisely as possible. ``Gold-plated'' lattice processes allow the determination of most CKM matrix elements, with the exception of $V_{tb}$.  These are simple processes with only one hadron in the initial state and at most one hadron in the final state, where the hadrons are stable (or at least narrow and far from threshold).  They are therefore the easiest to calculate with standard lattice QCD methods, and can now be obtained reliably from ``2+1'' flavor simulations that account for the dynamical $u$-, $d$-, and $s$-quarks with all sources of systematic error under control~\cite{Davies:2003ik}.  Figure~\ref{tab:CKM_decays} lists the gold-plated lattice quantities that can be used to obtain each CKM matrix element.  For a review of lattice calculations of most of these quantities, see the plenary talks in these proceedings by C. Aubin on heavy quark physics~\cite{Aubin:2009yh}, V. Lubicz on kaon physics~\cite{Lubicz_plenary}, and E. Scholz on light pseudoscalar meson masses and decay constants~\cite{Scholz:2009yz}.
\begin{figure}
\begin{center}
$\left( \begin{array}{ccc}
     		\mathbf{V_{ud}}  & \mathbf{V_{us}} & \mathbf{V_{ub}} \\*
		\pi \to \ell \nu  & K \to \ell \nu  & B \to \pi \ell \nu \\*
		 		& K \to \pi \ell \nu  &  \\*
     		\mathbf{V_{cd}} & \mathbf{V_{cs}} & \mathbf{V_{cb}}   \\*
		D \to \ell \nu  & D_s \to \ell \nu  & B \to D \ell \nu \\*
		D \to \pi \ell \nu  & D \to K \ell \nu  & B \to D^* \ell \nu \\*
		\mathbf{V_{td}}  & \mathbf{V_{ts}}  & \mathbf{V_{tb}}  \\*
		B_d \leftrightarrow \bar{B}_d  & B_s \leftrightarrow \bar{B}_s  & \\*
\end{array} \right)$
\caption{``Gold-plated'' processes on the lattice that can be used to obtain each CKM matrix element.  Neutral  $K^0 - \bar{K}^0 $ mixing is also a gold-plated process, and gives a constraint on the phase of the CKM matrix ($\rho, \eta$).}
\label{tab:CKM_decays}
\end{center}
\end{figure}

\begin{figure}[t]
\begin{center}
\includegraphics[width= 0.45\linewidth]{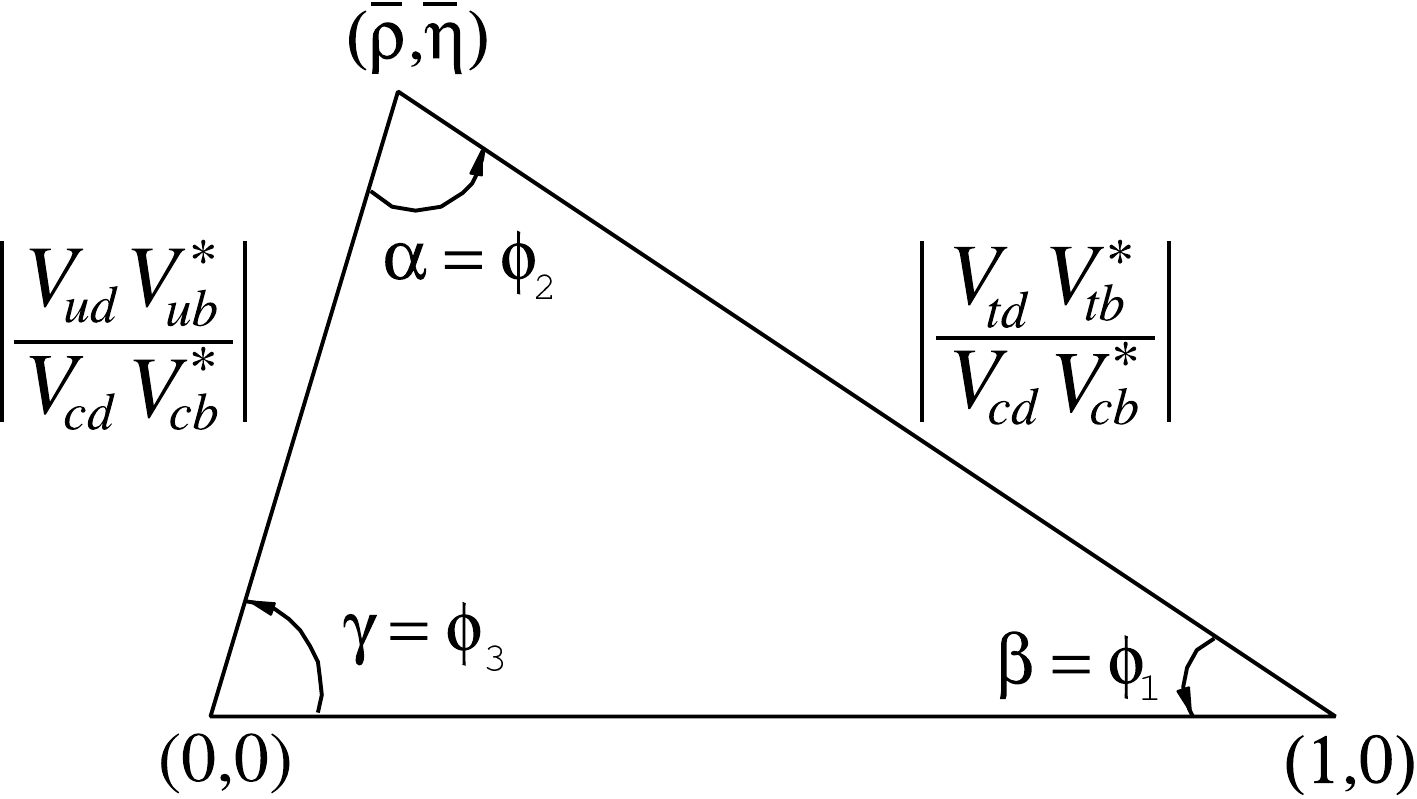}
\caption{The CKM unitarity triangle.  Figure from Ref.~\cite{Amsler:2008zzb}.}
\label{fig:UT_sketch}
\end{center}
\end{figure}

The quark mixing matrix is unitary, leading to normalization and orthogonality relations among the matrix elements.  In the case of three generations of Standard Model quarks, the most important relationship is
\begin{equation}
V_{ud} V_{ub}^* + V_{cd}V^*_{cb} + V_{td}V^*_{tb} = 0 \,.
\label{eq:UT_eq}
\end{equation}
Equation~(\ref{eq:UT_eq}) can be expressed as a triangle in the complex $\rho$-$\eta$ plane known as the CKM
unitarity triangle;  this is shown in Fig.~\ref{fig:UT_sketch}. 
Note that the CKM unitarity triangle is rescaled by $|V_{cd} V_{cb}^*|$ so that its base has unit length.  In practice, terms of $\CO(\lambda^4)$ and higher must be included in order to achieve sufficient accuracy for phenomenology.  Thus it is standard to work with the parameters $\bar\rho$ and $\bar\eta$, which are defined through all-orders in $\lambda$ as the apex of the triangle in Fig.~\ref{fig:UT_sketch}:
\begin{equation}
	\bar\rho + i \bar\eta = -\frac{V_{ub}^* V_{ud}}{V_{cb}^* V_{cd}} \,.
\end{equation}
Constraints on ($\bar\rho, \bar\eta$) can translated in a straightforward manner into constraints on original Wolfenstein parameters $(\rho, \eta)$.  

Deviations from the relationship in Eq.~(\ref{eq:UT_eq}) would indicate the presence of physics beyond the Standard Model.  It is likely that new physics will have additional quark flavor changing interactions and $CP$-violating phases.  Once theory and experiment are sufficiently precise, these will manifest themselves as
inconsistent determinations of the apex of the unitarity triangle ($\bar\rho, \bar\eta$).  One of the key goals in flavor physics is therefore to determine ($\bar\rho, \bar\eta$) as precisely as possible from as many independent processes as possible in order to search for new physics.  

Several important constraints on the CKM unitarity triangle require both experimental measurements and lattice QCD calculations of nonperturbative hadronic weak matrix elements.  Lattice calculations of neutral kaon mixing are needed to determine the mixing parameter $B_K$ and constrain the phase of the CKM matrix.  Computations of neutral $B$-meson decays and mixing are needed to obtain the decay constants $f_{B_q}$ and mixing parameters $B_{B_q}$ ($q=d,s$); these are used to determine $|V_{td}|$ and $|V_{ts}|$.  Calculations of the semileptonic $B$-meson decay $B \to \pi \ell \nu$ are needed to compute the form factor $f_+(q^2)$, which allows a determination of $|V_{ub}|$, while those of the decays $B \to D \ell \nu$ and $B \to D^* \ell \nu$ are needed to compute the form factors $F(1), G(1)$, which allow one to obtain $|V_{cb}|$.

\section{Status of the CKM matrix}
\label{sec:CKM_status}

In this section I focus on tests of the Standard Model CKM framework, emphasizing the role played by lattice QCD inputs.  In Sec.~\ref{sec:1st_row} I review the status of first-row unitarity, while in Sec.~\ref{sec:CKM_UT} I review that of the global CKM unitarity triangle fit. 

\subsection{First-row unitarity}
\label{sec:1st_row}

In the Standard Model, elements of the first row of the CKM matrix must obey the following relation:
\begin{equation}
	|V_{ud}|^2 + |V_{us}|^2 + |V_{ub}|^2 = 1 \,. \label{eq:1st_row}
\end{equation}
Because $|V_{ub}| \sim \CO(10^{-3})$, Eq.~(\ref{eq:1st_row}) is essentially a constraint on the relationship between $|V_{ud}|$ and $|V_{us}|$.  A significant deviation in this relationship would be evidence for new physics.  

The CKM matrix elements $|V_{ud}|$ and $|V_{us}|$ are known to sub-percent accuracy~\cite{Antonelli:2008jg}:
\begin{eqnarray}
	|V_{ud}| &=& 0.97418 \pm 0.00026 , \\
	|V_{us}| &=& 0.2246 \pm 0.0012 , \label{eq:Vud}
\end{eqnarray}
where $|V_{ud}|$ is obtained from nuclear $\beta$-decays~\cite{Towner:2007np} and $|V_{us}|$ relies on a $N_f = 2+1$ flavor lattice QCD calculation of the $K \to \pi \ell \nu$ form factor by the RBC and UKQCD collaborations~\cite{Antonio:2007mh}.
The ratio $|V_{us}|/|V_{ud}|$ is also known to this accuracy~\cite{Antonelli:2008jg}:
\begin{equation}
	|V_{us}|/|V_{ud}| = 0.2321 \pm 0015 , \label{eq:Vus/Vud}
\end{equation}
which uses a $N_f = 2+1$ flavor lattice QCD determination of $f_K/f_\pi$ by the HPQCD collaboration~\cite{Follana:2007uv}.  Equation~(\ref{eq:1st_row}) therefore allows for a precision test of the Standard Model and probe of new physics.

\begin{figure}[t]
\begin{center}
\includegraphics[width= 0.485\linewidth]{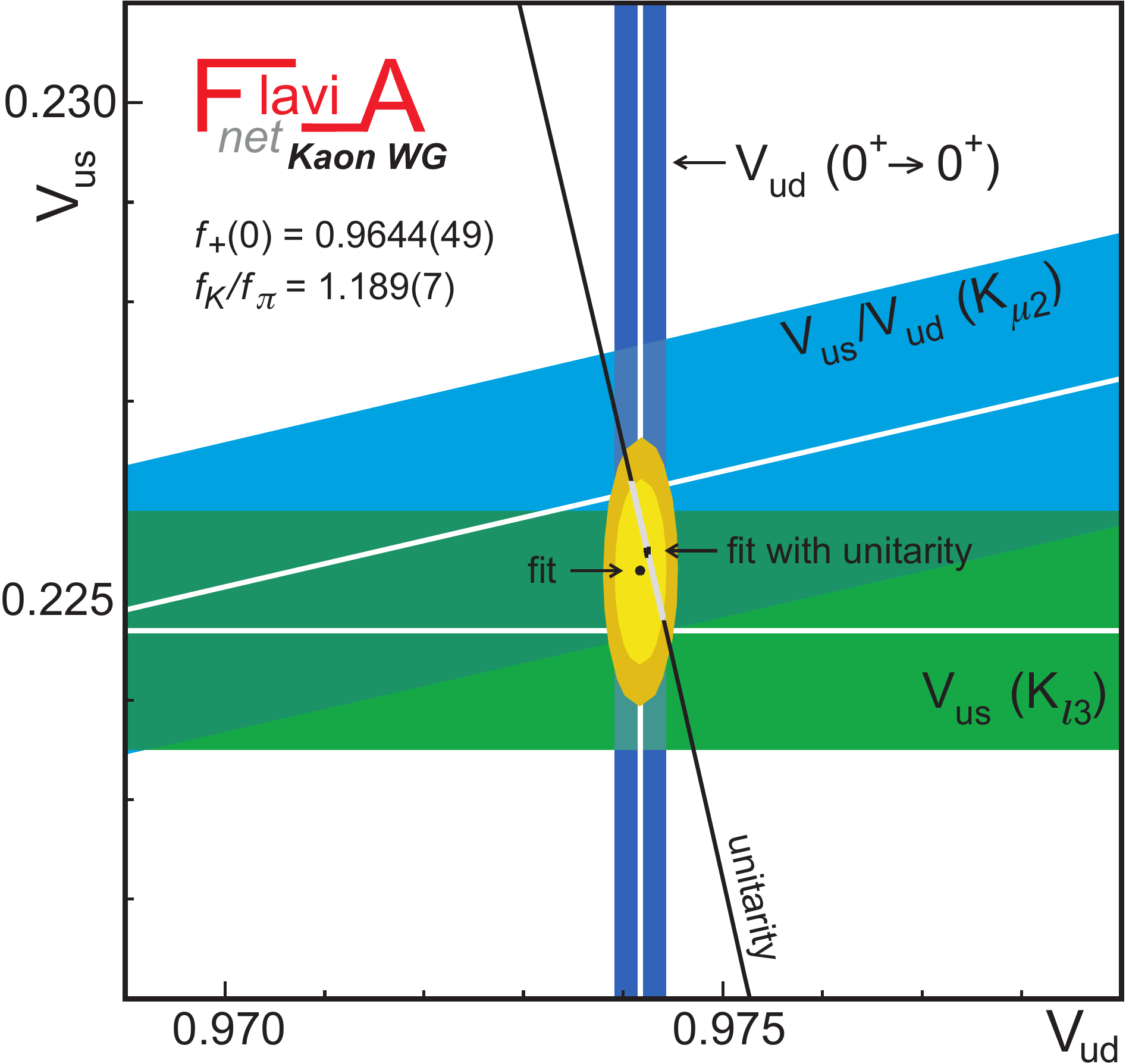}
\caption{Combined fit to $|V_{ud}|$, $|V_{us}|$, and $|V_{us}|/|V_{ud}|$ both with and without the constraint from first-row unitarity~\cite{Antonelli:2008jg}.  The fit with first-row unitarity imposed has a $\chi^2/{\rm d.o.f.} = 0.65$ and ${\rm C.L.} = 42\%$.}
\label{fig:1st_row}
\end{center}
\end{figure}

The Flavianet collaboration has performed a combined fit to $|V_{ud}|$, $|V_{us}|$, and $|V_{us}|/|V_{ud}|$ using the values in Eqs.~(\ref{eq:Vud}) and~(\ref{eq:Vus/Vud}), both with and without imposing the constraint from first-row unitarity~\cite{Antonelli:2008jg}.  This is shown in Fig.~\ref{fig:1st_row}.
Without assuming unitarity, the result of the combined fit is
\begin{eqnarray}
	|V_{ud}| = 0.97417(26)\,, \\
	|V_{us}| = 0.2253(9) \,,
\end{eqnarray}
which is consistent with first-row unitarity at subpercent level:
\begin{equation}
	|V_{ud}|^2 + |V_{us}|^2 + |V_{ub}|^2 = 0.9998(6) \,.
\end{equation}
In addition, the fit including the unitarity constraint has a confidence level of 67\%.  Thus there is no indication of new physics in the first row of the CKM unitarity triangle given the current levels of experimental and theoretical precision.

\subsection{The CKM unitarity triangle}
\label{sec:CKM_UT}

In this section I summarize the current status of the global fit of the CKM unitarity triangle.  For this discussion, I use the recent averages of the necessary lattice QCD inputs provided in Ref.~\cite{Laiho:2009eu}; these are given in Table~\ref{tab:LQCD_inputs}.  
\begin{table}
\begin{center}
\begin{tabular}{lr}
\hline\hline
Quantity & Error \\\hline
$\hat B_K = 0.725 \pm 0.026$ & $\sim$ 4\%\\
$f_{B_s} \sqrt{\hat{B}_{B_s}} = 275 \pm 13$ & $\sim$ 5\% \\
$\xi = 1.243 \pm 0.028 $ & $\sim$ 2\%\\
$|V_{ub}|_\textrm{excl.} = ( 3.42 \pm 0.37) \times 10^{-3} $ & $\sim$ 11\% \\
$|V_{cb}|_\textrm{excl.} = ( 38.6 \pm 1.2) \times 10^{-3}$ & $\sim$ 3\%\\
$\kappa_\varepsilon = 0.92 \pm 0.01$ & $\sim$ 1\%\\
$f_K = 155.8 \pm 1.7$ MeV & $\sim$ 1\% \\
\hline\hline
\end{tabular}
\end{center}
\caption{Current status of lattice inputs to the global fit of the CKM unitarity triangle~\cite{Laiho:2009eu}.  These were obtained by averaging all available $N_f = 2+1$ results documented in proceedings and publications that contain complete error budgets, and account for correlations between different calculations in a conservative manner.}
\label{tab:LQCD_inputs}
\end{table}
These results were obtained by averaging all available $N_f = 2+1$ flavor results documented in proceedings in publications that contain complete error budgets, and therefore do not include any new results as of Lattice 2009.  
See the plenary talks of C. Aubin~\cite{Aubin:2009yh}, V. Lubicz~\cite{Lubicz_plenary}, and E. Scholz~\cite{Scholz:2009yz} in these proceedings for discussions of more recent preliminary and $N_f=2$ results.  
For the computation of the lattice QCD averages in Ref.~\cite{Laiho:2009eu}, asymmetric errors are symmetrized and it is assumed that errors are normally distributed.  Furthermore, when there are correlations between a source of error in different lattice calculations, it is conservatively assumed that the correlation is 100\%.

\subsubsection{$|V_{cb}|$ and $|V_{ub}|$}

Tensions exist between the inclusive and exclusive determinations of both $|V_{cb}|$ and $|V_{ub}|$.  The inclusive values are independent of lattice QCD, while the exclusive results rely on lattice computations of $B$-meson semileptonic form factors.

\begin{figure}[t]
\begin{center}
\includegraphics[height= 0.23\linewidth]{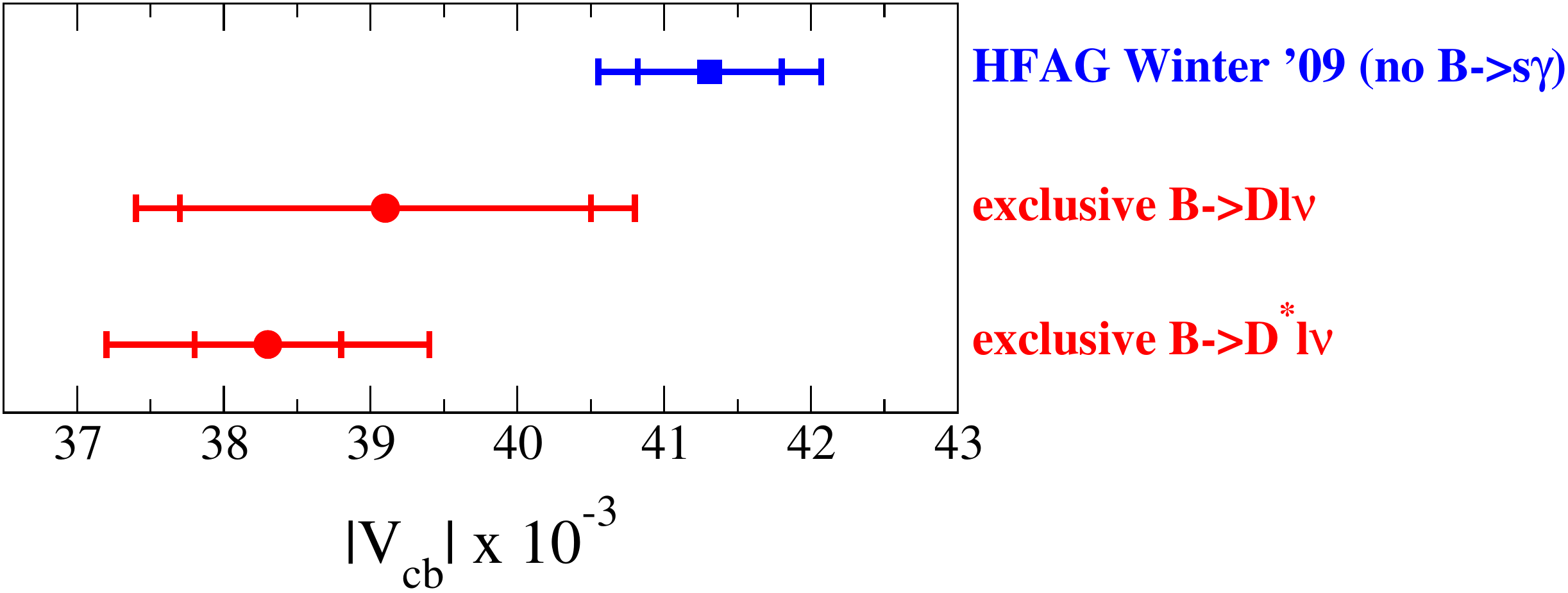}
\caption{Tension between inclusive (blue square) and exclusive (red circles) determinations of $|V_{cb}|$.  The inner and outer errors bars on the inclusive value denote the errors from the moment fit and from other systematics, respectively.  The inner and outer errors on the exclusive values are experimental and theoretical, respectively.   The inclusive determination~\cite{HFAG_Wi09} disagrees with the exclusive determination from $B\to D\ell\nu$ by $1.2\sigma$ and from $B\to D^*\ell\nu$ by $2.3\sigma$.}
\label{fig:Vcb}
\end{center}
\end{figure}

Figure~\ref{fig:Vcb} compares the inclusive determination of $|V_{cb}|$~\cite{HFAG_Wi09} with exclusive determinations based on lattice QCD calculations of the $B \to D \ell \nu$~\cite{Okamoto:2004xg} and $B \to D^* \ell \nu$~\cite{Bernard:2008dn} form factors at zero recoil by the Fermilab Lattice and MILC collaborations.  
The tension between inclusive and exclusive values is $\sim 1.2$--$2.3\sigma$.  This discrepancy is significant because $|V_{cb}|$ is used to normalize the base of the unitarity triangle.  In fact, the error in $|V_{cb}|$ is now the limiting uncertainty in some of the unitarity triangle constraints, as shown in the next subsection.  Therefore improving the errors in exclusive $|V_{cb}|$ should be a high priority for lattice QCD calculations.

\begin{figure}[t]
\begin{center}
\includegraphics[height= 0.23\linewidth]{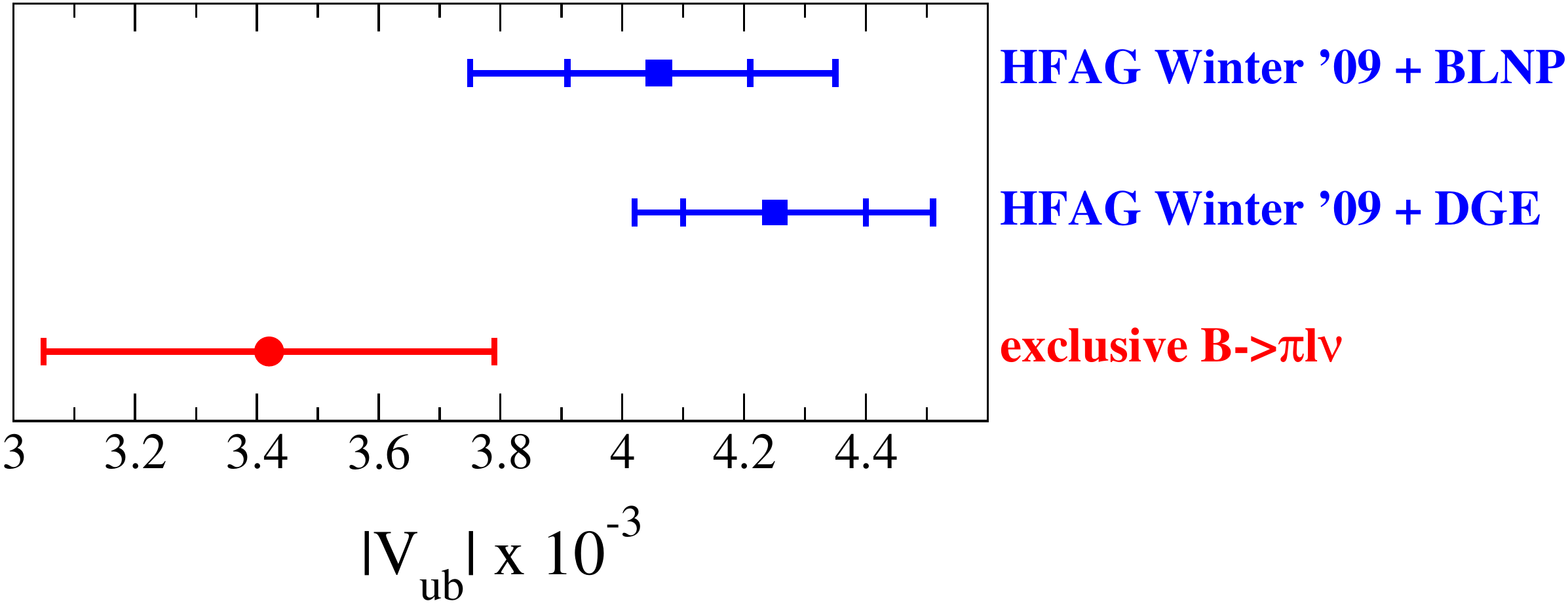}
\caption{Tension between inclusive (blue squares) and exclusive (red circle) determinations of $|V_{ub}|$.  The inner and outer errors on the inclusive values are experimental and theoretical, respectively.  The error bar on the exclusive value denotes the total statistical plus systematic error added in quadrature. The exclusive determination disagrees with the inclusive determinations~\cite{HFAG_Wi09} by $1$--$2\sigma$ depending upon the theoretical framework used to obtain the inclusive value.}
\label{fig:Vub}
\end{center}
\end{figure}

Figure~\ref{fig:Vub} compares two representative inclusive determinations of $|V_{ub}|$~\cite{HFAG_Wi09} with the exclusive determination based on lattice QCD calculations of the $B \to \pi \ell \nu$ form factor by the HPQCD collaboration~\cite{Dalgic:2006dt} and by the Fermilab Lattice and MILC collaborations~\cite{Bailey:2008wp}.
The $1$--$2\sigma$ tension between inclusive and exclusive determinations of $|V_{ub}|$ is less worrisome, however, than in the case of $|V_{cb}|$ because the inclusive value of $|V_{ub}|$ varies significantly depending upon theoretical framework, and is highly sensitive to the input value of the $b$-quark mass~\cite{Gambino:2008zz}.  Nevertheless, if the discrepancy in $|V_{ub}|$ holds up as the size of the theoretical errors are reduced, it could be a sign of non-Standard Model $(V+A)$ currents.  A right-handed $b\to u$ current would affect the extraction of $|V_{ub}|$ from $B\to \pi \ell \nu$ semileptonic decay differently than the extraction of $|V_{ub}|$ from $B \to X_u \ell \nu$ inclusive semileptonic decay because $B\to \pi \ell \nu$ decay proceeds through only the vector part of the left-handed current, whereas $B \to X_u \ell \nu$ inclusive decay is the sum of several exclusive channels that proceed via different linear combinations of the vector and axial-vector currents.  The right-handed current hypothesis could be tested using $|V_{ub}|$ determined from exclusive $B \to \rho \ell \nu$ semileptonic decay since $B \to \rho \ell \nu$ proceeds through the entire left-handed current.  One could use the values of $|V_{ub}|$ from $B\to \pi \ell \nu$ and $B \to X_u \ell \nu$ to make a prediction for the size of the non-Standard Model $(V+A)$ current, and make a prediction for the value of $|V_{ub}|$ from $B \to \rho \ell \nu$ given this hypothesis.  This motivates the need for a reliable lattice QCD calculation of the $B \to \rho \ell \nu$ form factors.

\subsubsection{Neutral kaon mixing}

The amount of direct $CP$-violation in the neutral kaon system, $\varepsilon_K$, which has been measured experimentally to sub-percent precision, constrains the apex of the CKM unitarity triangle:
\begin{equation}
\small{|\varepsilon_K|  = \kappa_\varepsilon C_\varepsilon B_K A^2 \bar\eta\{-\eta_1 S_0 (x_c)(1-\lambda^2/2) + \eta_3 S_0(x_c,x_t) + \eta_2 S_0 (x_t) A^2 \lambda^2 (1- \bar\rho)\}} .
\label{eq:epsK}
\end{equation}
Until recently, the uncertainty in the unitarity triangle constraint from $\varepsilon_K$ was primarily due to the uncertainty in lattice QCD calculations of the hadronic matrix element $B_K$.  This is no longer true, however, thanks to the $N_f = 2+1$ calculations by the RBC and UKQCD collaborations~\cite{Antonio:2007pb} and by Aubin, Laiho, and Van de Water~\cite{Aubin:2009jh}, which constrain $B_K$ to $\sim (4-6)\%$ accuracy.  Furthermore, the errors in $B_K$ will continue to decrease as the RBC and UKQCD collaborations update their analysis using two lattice spacings~\cite{Kelly:2009fp} and as independent determinations using twisted-mass fermions~\cite{Bertone:2009bu} and improved staggered fermions~\cite{Bae:2009tf,Kim:2009te,Yoon:2009th,Kim:2009ti} become available.

\begin{figure}[t]
\begin{center}
\includegraphics[width= 0.55\linewidth]{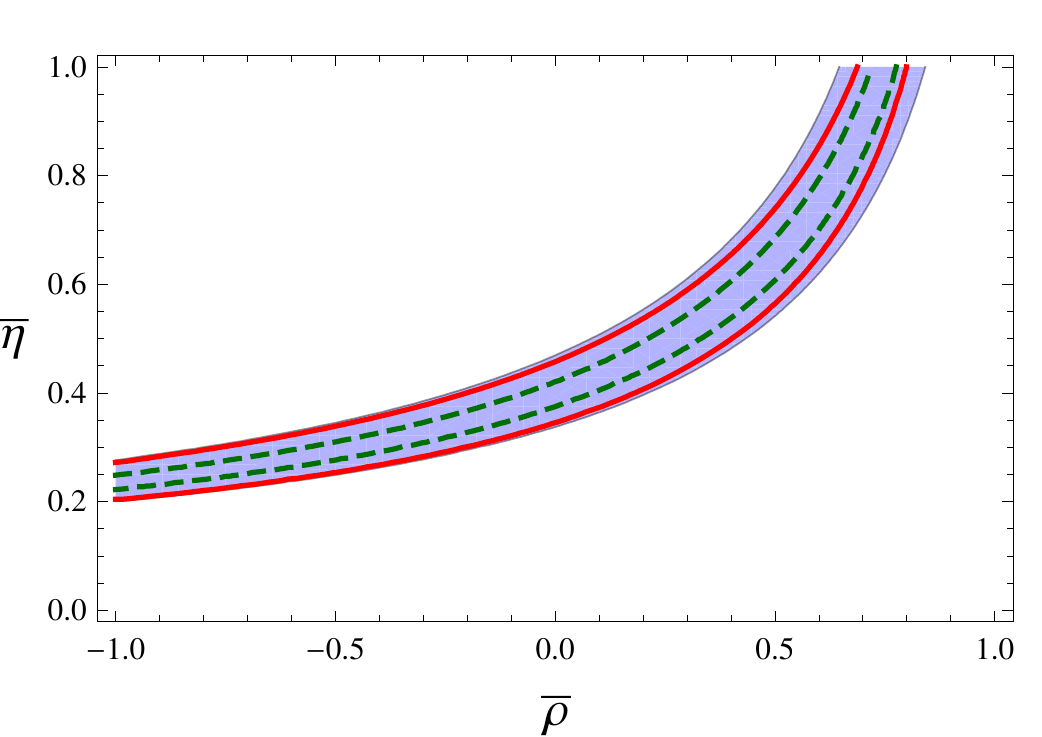}
\caption{Contributions of $|V_{cb}|$ (solid red line) and $\hat{B}_K$ (dashed green line) to the uncertainty in the $\varepsilon_K$ band. The errors introduced by the remaining inputs to the $\varepsilon_K$ band are negligible.  Figure from Ref.~\cite{Laiho:2009eu}.}
\label{fig:EpsK}
\end{center}
\end{figure}

The dominant error in the $\varepsilon_K$ band is now from the Wolfenstein parameter $A$, which is proportional to the CKM matrix element $|V_{cb}|$.  Although $|V_{cb}|$ is known to $\sim 2\%$ accuracy, it enters Eq.~(\ref{eq:epsK}) as the fourth power and therefore contributes an $\sim 8\%$ error to the $\varepsilon_K$ band.  Figure~\ref{fig:EpsK} illustrates this point by showing the relative sizes of the contributions from $B_K$ and $|V_{cb}|$ to the $\varepsilon_K$ band. 
It is easy to see that further improvements in $B_K$ will not tighten the constraint from $\varepsilon_K$ on the apex of the CKM unitarity triangle without an improvement in $|V_{cb}|$.

Given the recent reductions in the errors on $B_K$ and $|V_{cb}|$, it is now important to include $5 - 10\%$ corrections to Eq.~(\ref{eq:epsK}) that have been omitted from most previous unitarity triangle analyses.    In the Standard Model, direct $CP$-violation is parameterized as:
\begin{equation}
\varepsilon_K = e^{i\phi_\varepsilon}\sin{\phi_\varepsilon}\left(\frac{\textrm{Im}(M^K_{12})}{\Delta M_K}+P_0 \right) ,
\label{eq:epsK_full}
\end{equation}
where the first term in parentheses is the short-distance contribution which is proportional to $B_K$ and the second term, $P_0$, is the long-distance contribution which is related to the ratio of $K \to \pi \pi$ decay amplitudes in the $\Delta I = 1/2$ channel.  Because the phase $\phi_\varepsilon$ is close to  $45^\circ$ and the size of $P_0$ is small compared to the short-distance contribution, it is common to set $\phi_\varepsilon = 45^\circ$ and $P_0 = 0$; this leads to the standard expression given in Eq.~(\ref{eq:epsK}).  Buras and Guadagnoli recently pointed out, however, that the corrections due to $\phi_\varepsilon \neq 45^\circ$ and $P_0 \neq 0$, are no longer negligible given the sizes of other errors~\cite{Buras:2008nn}.  

The effects of $\phi_\varepsilon \neq 45^\circ$ and $P_0 \neq 0$ can be parameterized by an overall multiplicative factor~\cite{Anikeev:2001rk}:
\begin{equation}
\varepsilon_K \approx \kappa_\varepsilon \frac{\textrm{Im}(M^K_{12})}{\Delta M_K} ,
\end{equation}
where $\kappa_\varepsilon$ is given by
\begin{equation}
\kappa_\varepsilon = \sqrt{2} \sin{\phi_\varepsilon} \left( 1-\frac{1}{\omega}\textrm{Re}(\varepsilon'_K/\varepsilon_K) + \frac{\textrm{Im} A_2 / \textrm{Re} A_2}{\sqrt{2}|\varepsilon_K|} \right) .
\label{eq:kap_eps}
\end{equation}
Fortunately, assuming the Standard Model, the majority of inputs to $\kappa_\varepsilon$ are well-known from experiment.  The remaining unknown, $\textrm{Im} A_2$, can be obtained from lattice QCD calculations of $K \to \pi \pi$ matrix elements in the $\Delta I = 1/2$ channel.  The only $N_f = 2 +1$ flavor determination of this quantity is by the RBC and UKQCD collaborations~\cite{Li:2008kc}, and has rather large systematic errors associated with the use of leading-order chiral perturbation theory.  Nevertheless, the contribution to Eq.~(\ref{eq:kap_eps})  from $\textrm{Im} A_2$ is small and leads to only a $\sim 1\%$ uncertainty in $\kappa_\varepsilon$~\cite{Laiho:2009eu}:
\begin{equation}
	\kappa_\varepsilon = 0.92 \pm 0.01 \,.
\end{equation}
This result agrees with the estimate of Buras and Guadagnoli~\cite{Buras:2008nn} and lowers the SM prediction for $\varepsilon_K$ by 8\%.  The correction factor $\kappa_\varepsilon$ is included in the global unitarity triangle fits presented in the following section~\cite{Laiho:2009eu}, and has also recently been included by the UTfit collaboration~\cite{Bona:2009ze}.  It has not yet been implemented by the CKMfitter group~\cite{CKMfitter}.

\subsubsection{Global fit of the CKM unitarity triangle}

\begin{figure}[t]
\begin{center}
\includegraphics[width= 0.56\linewidth]{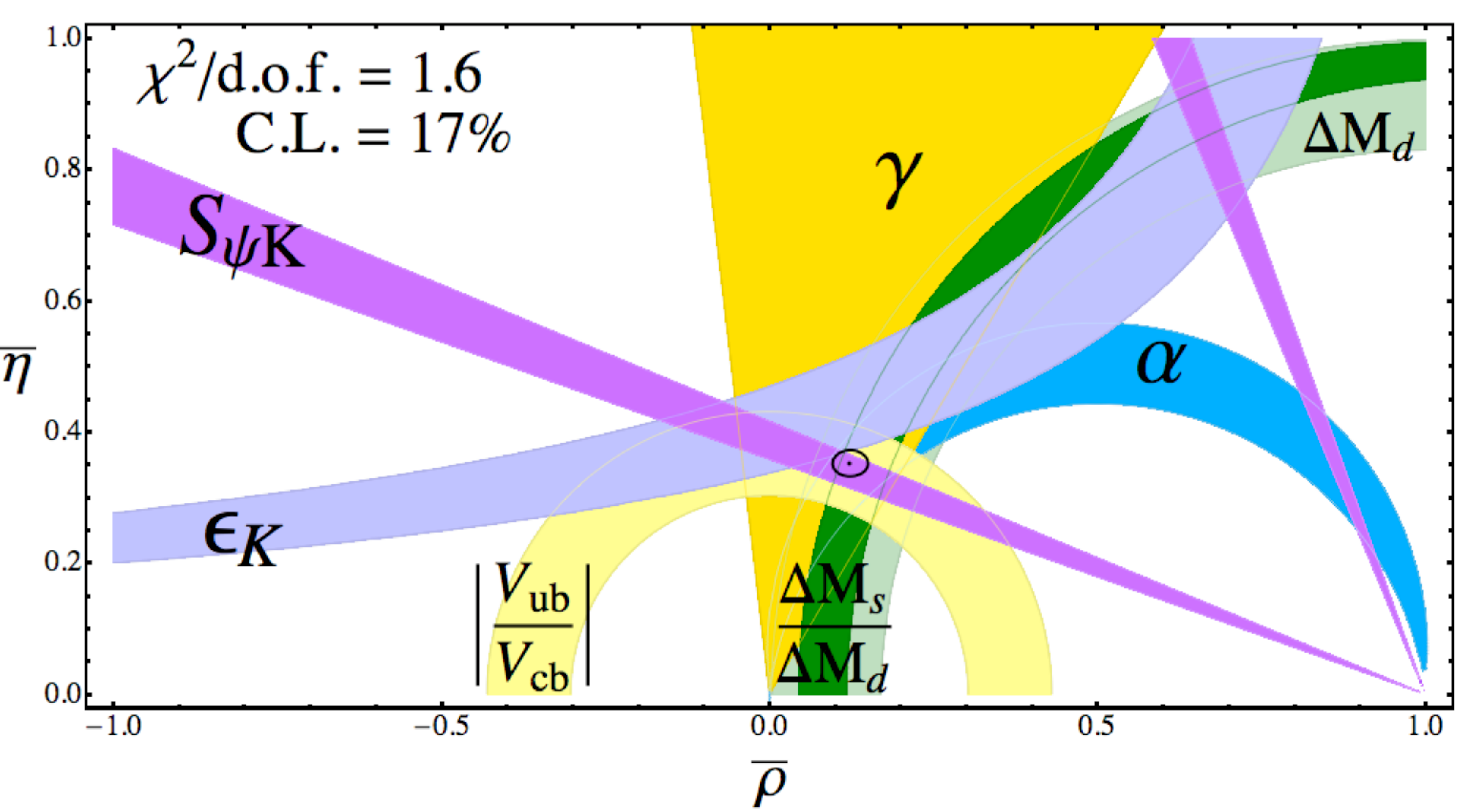}
\caption{Global fit of the CKM unitarity triangle~\cite{Laiho:2009eu} .  The current fit is consistent with the Standard Model at the 23\% level.  The constraints from $\varepsilon_K$, $|V_{ub}|/|V_{cb|}$, $\Delta M_s / \Delta M_d$, and $\Delta M_d$ are all limited by theoretical uncertainties from lattice QCD.}
\label{fig:UTfit}
\end{center}
\end{figure}

Figure~\ref{fig:UTfit} shows the current status of the global CKM unitarity triangle fit using the lattice QCD inputs presented in Table~\ref{tab:LQCD_inputs}~\cite{Laiho:2009eu} .
Although the average of inclusive and exclusive determinations of $|V_{cb}|$ is used, the error in the average is inflated in order to account for the inconsistency between the two values following the prescription of the Particle Data Group~\cite{Amsler:2008zzb}.  Only the exclusive determination of $|V_{ub}|$ is used, however, because the inclusive determination varies so much depending on the theoretical framework.  The confidence level of the global fit is 17\%; thus current observations are consistent with Standard Model expectations given the present level of theoretical precision.

\begin{figure}[t]
\begin{center}
\includegraphics[width= 0.56\linewidth]{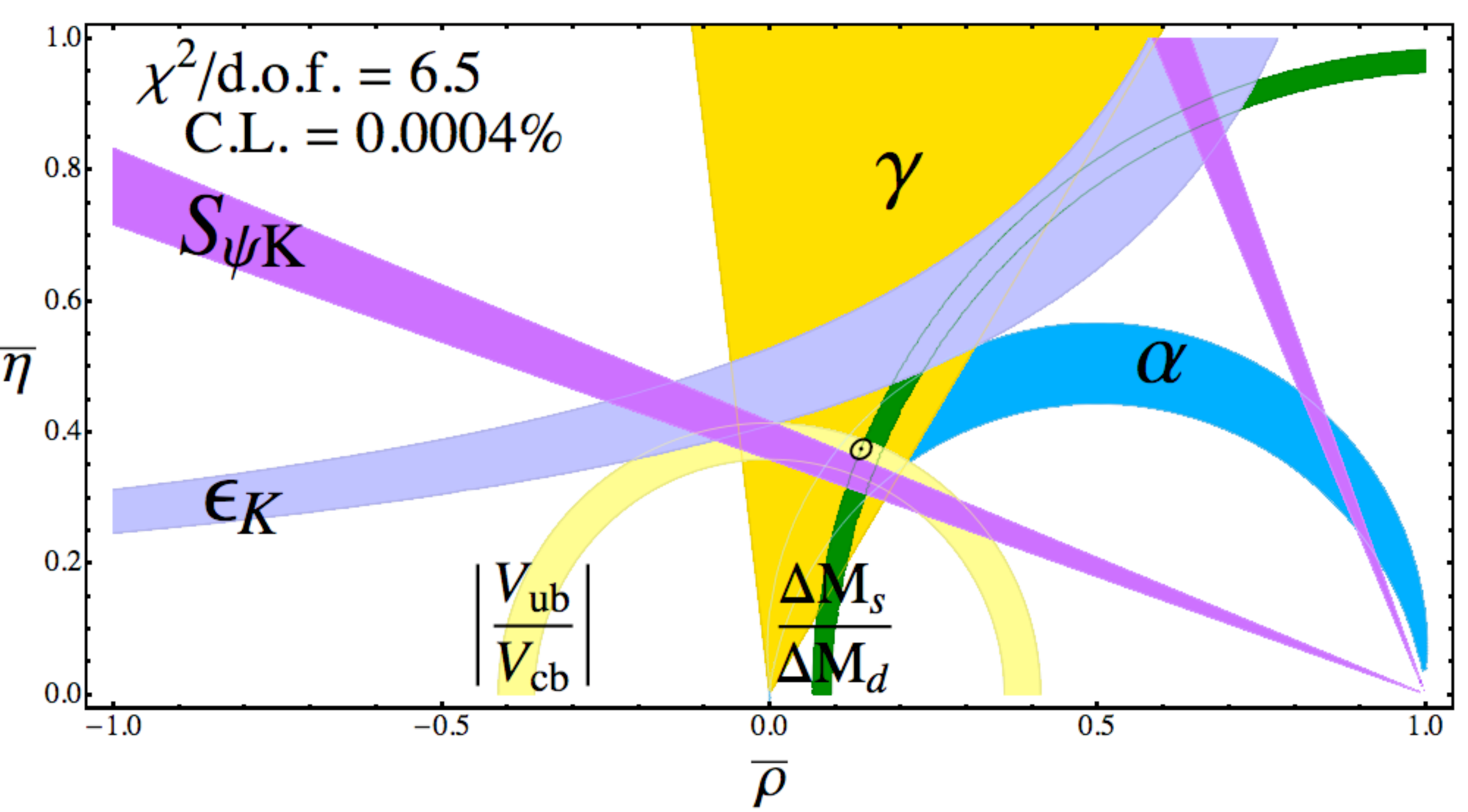}
\caption{Potential impact of future lattice determinations on the global unitarity triangle fit.  If the theoretical errors in all of the lattice QCD inputs are reduced to 1\% with the central values fixed, the fit would no longer be consistent with Standard Model expectations.  Figure courtesy of E. Lunghi.}
\label{fig:UTfit_future}
\end{center}
\end{figure}

Currently the constraints from $\varepsilon_K$, $\Delta m_s / \Delta m_d$, and $|V_{ub}/V_{cb}|$ are limited by uncertainties in the lattice QCD calculations of $|V_{cb}|_{\rm excl.}$, $\xi$, and $|V_{ub}|_{\rm excl.}$, respectively.  In order to show the potential impact of future lattice calculations, it is therefore an interesting exercise to repeat the global CKM unitarity triangle fit after reducing the lattice uncertainties in $\xi$, $B_K$, $|V_{cb}|_{\rm excl.}$, and $|V_{ub}|_{\rm excl.}$ to 1\% (with central values fixed).   The resulting fit is shown in Fig.~\ref{fig:UTfit_future}.
In this case, only the exclusive determination of $|V_{cb}|$ is used because combining it with the inclusive determination becomes problematic when the lattice errors are so small.  The confidence level of this hypothetical fit is 0.0004\%, indicating that this scenario would no longer be compatible with the Standard Model.  Thus improved lattice QCD calculations of hadronic weak matrix elements could allow the observation of new physics in the quark flavor sector with a high significance.

\section{Hints of new physics in the flavor sector}
\label{sec:NP_hints}

Although most observations in the flavor sector are consistent with Standard Model expectations, there are currently several 2--3$\sigma$ tensions that may indicate the presence of new physics.  In the following subsections I present those hints of new physics that rely on lattice QCD calculations of weak matrix elements.  It is worthwhile, however, briefly mentioning first two others that do not require lattice inputs.  In the case of all tensions, we must, of course, wait and see whether their significance increases or decreases with improved experimental and theoretical precision.  

The $B_s$-mixing phase in the Standard Model is given by $\beta_s^\textrm{SM} = \textrm{arg}(-V_{ts}V_{tb}^*/V_{cs}V_{cb}^*) \approx 0.02$.  This prediction disagrees, however, with the measured world average based on flavor-tagged analyses of $B_s \to J/\psi \phi$ decays by 2.2$\sigma$~\cite{HFAG_BsPhase}.  If this discrepancy remains as the experimental errors are reduced, it would indicate the presence of new sources of $CP$-violation beyond the phase of the CKM matrix.  Such new phases would lead to correlated effects between $\Delta B=2$ processes and $b\to s$ decays, thereby making improved measurements of $CP$-violation in $b \to s$ penguin decays particularly important~\cite{Bona:2008jn}.  Thus is is interesting that the ``effective'' value of the angle $\beta$ in the CKM unitarity triangle obtained from $b \to q \bar{q} s$ penguin decays is lower than the value of $\beta$ obtained from tree-level $b \to c \bar{c} s$ decays, which are expected to be less sensitive to new physics.  For example, $\sin(2\beta)_\textrm{eff}$ determined from the penguin decay $B \to \phi K^0$ is $\sim 1.3 \sigma$ lower than the average $\sin(2\beta)$ from tree-level decays~\cite{HFAG_beta_penguin}.   Although this discrepancy is not statistically significant, the value of $\sin(2\beta)_\textrm{eff}$ obtained from various penguin decay modes is systematically lower than the average $\sin(2\beta)$ from tree-level decays in almost all decay channels.  Addressing these puzzles will require improved determinations of the $B_s$-mixing phase and of $\sin(2\beta)_\textrm{eff}$ from $b \to q \bar{q} s$ penguin decays at LHCb and the super-$B$ factory.  


\subsection{Tension in the CKM unitarity triangle}

\begin{figure}[t]
\begin{center}
\includegraphics[width= 0.48\linewidth]{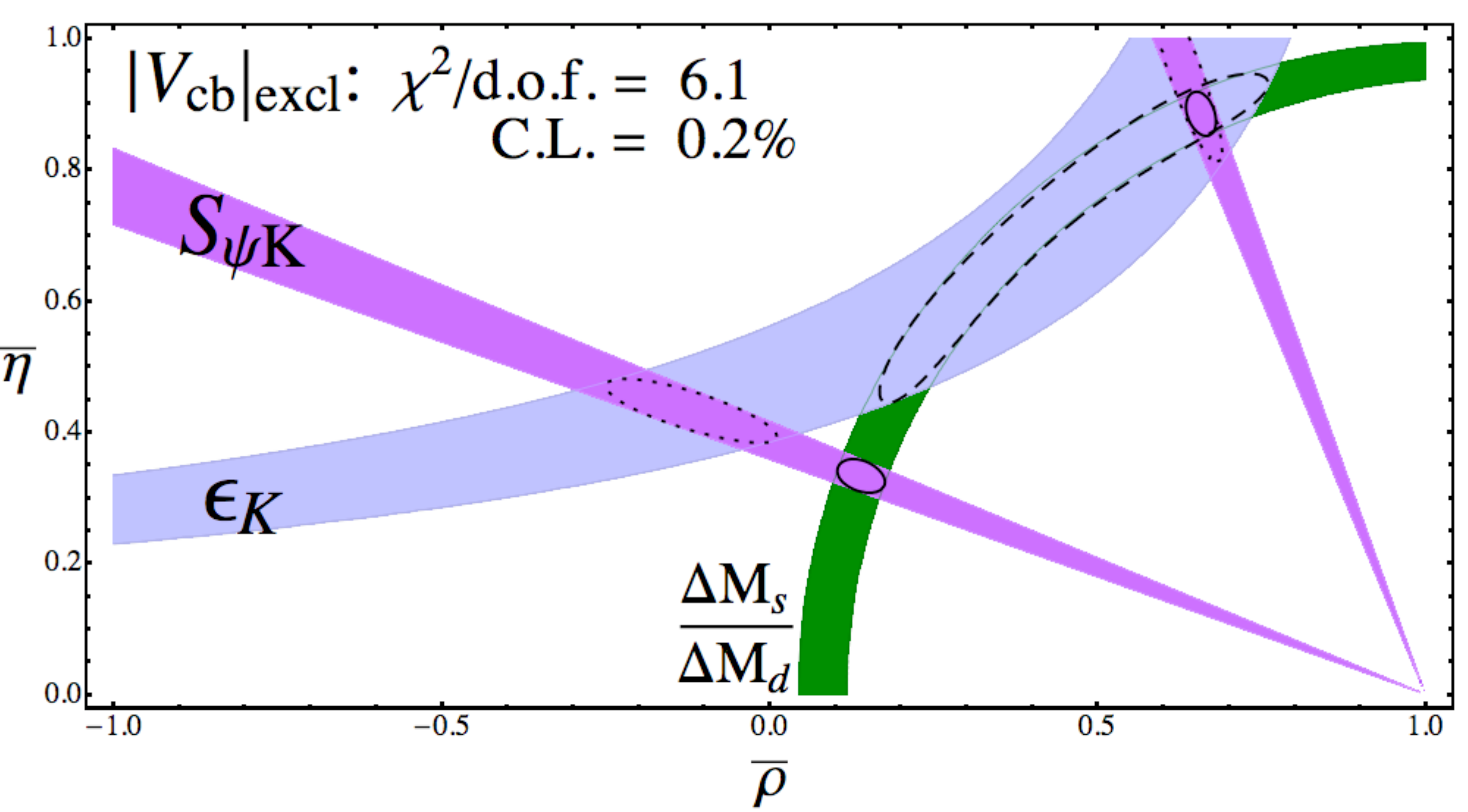}
\includegraphics[width= 0.48\linewidth]{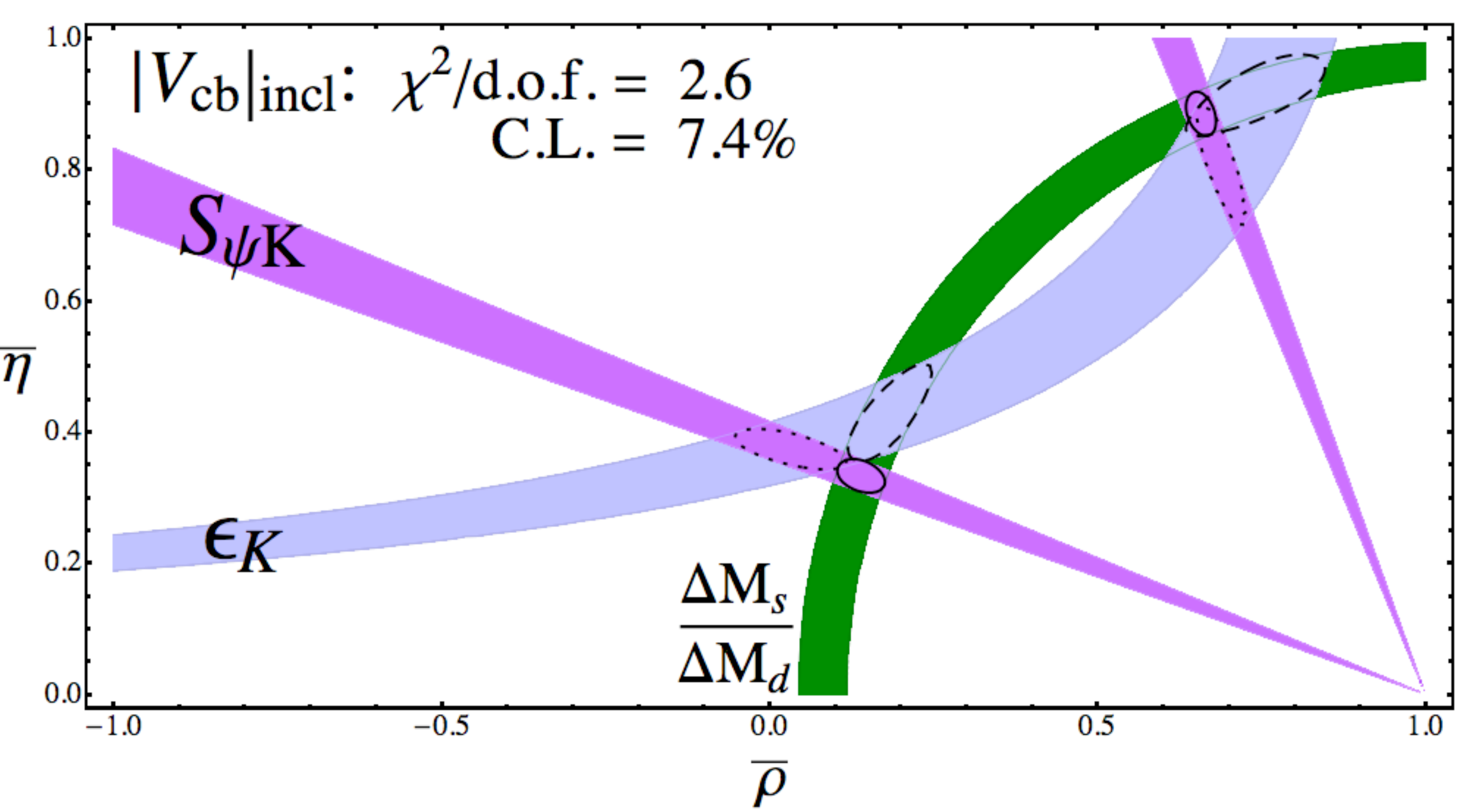}
\caption{Tension in the CKM unitarity triangle~\cite{Laiho:2009eu}. The confidence levels shown in the figures correspond to the unitarity triangle fit including all three constraints, while the solid, dashed and dotted contours are obtained by omitting $\varepsilon_K$,  $S_{\psi K}$ and $\Delta M_{B_s}/\Delta M_{B_d}$, respectively. The left and right panels use the exclusive and inclusive determinations of $|V_{cb}|$.}
\label{fig:UT_tension}
\end{center}
\end{figure}

As noted in the introduction, there is a 2--3$\sigma$ tension in the CKM unitarity triangle~\cite{Lunghi:2008aa,Bona:2009ze,Laiho:2009eu}.  If the constraints on $(\bar\rho, \bar\eta)$ from $\alpha$, $\gamma$, and $|V_{ub}|$ are omitted, it is easy to see that the tension is really between the three most precise constraints, which are those from $\varepsilon_K$, $\Delta m_s / \Delta m_d$, and $\sin (2 \beta)$~\cite{Lunghi:2008aa}; this is illustrated in Fig.~\ref{fig:UT_tension}.  
Furthermore, the degree of tension is extremely sensitive to the value of $|V_{cb}|$.  The confidence level of a fit to $\varepsilon_K$, $\Delta m_s / \Delta m_d$, and $\sin (2 \beta)$ using the exclusive determination of $|V_{cb}|$ is only 0.2\% (left plot of Fig.~\ref{fig:UT_tension}), whereas the confidence level using the inclusive determination is almost 50 times greater at 8.9\% (right plot of Fig.~\ref{fig:UT_tension}).

One way to quantify the significance of the tension is to leave one input as a
free parameter in the unitarity triangle fit and make a prediction for it based on the remaining constraints.
For example, the value of $|V_{cb}|$ preferred by $\varepsilon_K$,  $\sin(2 \beta)$ and $\Delta M_{B_s}/\Delta M_{B_d}$ is~\cite{Laiho:2009eu}:
\begin{equation}
	|V_{cb}|_{\rm fit} = (43.0 \pm 0.9) \times 10^{-3} \,.
\end{equation}
This deviates by$3.0\sigma$ and $1.3\sigma$ from the exclusive and inclusive determinations of $|V_{cb}|$, respectively.  Because of the sensitivity of the unitarity triangle analysis to the value of $|V_{cb}|$ (and the discrepancy between inclusive and exclusive determinations), obtaining $|V_{cb}|$ should be a high priority for lattice QCD calculations.  

Alternatively, the value of $B_K$ preferred by the other inputs depends upon the value of $|V_{cb}|$~\cite{Laiho:2009eu}: 
\begin{eqnarray}
( \hat B_K )_{\rm fit}  =  
\begin{cases}
1.09 \pm 0.12 & \left|V_{cb} \right|_{\rm excl} \cr 
0.903 \pm 0.086 &\left|V_{cb} \right|_{\rm incl} \cr 
0.98 \pm 0.10 & \left|V_{cb} \right|_{\rm excl+incl} \cr 
\end{cases}
\end{eqnarray}
These deviate from the lattice QCD average for $B_K$ at the $2.9\sigma$, $2.0\sigma$ and $2.4\sigma$ levels, respectively.  
The observed tension between the $\varepsilon_K$ band and the other unitarity triangle constraints relies on the inclusion of the correction factor $\kappa_\varepsilon$, which raises location of the $\varepsilon_K$ band in the $\bar{\rho}$-$\bar{\eta}$ plane by 8\%.

\subsection{The ``$f_{D_s}$ puzzle''}

\begin{figure}[t]
\begin{center}
\psfrag{200}[b][b][0.9]{$\to$}
\includegraphics[width= 0.6\linewidth]{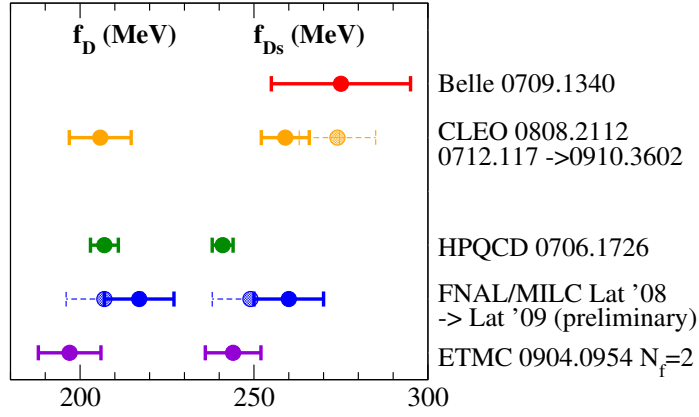}
\caption{Comparison of experimental measurements and lattice QCD calculations of $f_D$ and $f_{D_s}$.  The majority of the results shown agree, but the HPQCD calculation of $f_{D_s}$ disagrees with the CLEO measurement by $2.4 \sigma$.  This discrepancy has yet to be confirmed, however, by an independent lattice calculation with comparably small theoretical errors.  Furthermore, the HPQCD determination of $f_{D_s}$ will likely increase by $\sim 1.5 \sigma$ when the recently updated value of the scale $r_1$~\cite{McNeile:2009eq} is taken into account.}
\label{fig:fDs}
\end{center}
\end{figure}

The HPQCD collaboration's published lattice QCD calculation of the leptonic decay constant $f_{D_s}$~\cite{Follana:2007uv} disagrees with the CLEO experimental measurement~\cite{Collaboration:2009tk} by 2.4$\sigma$, despite the fact that HPQCD's determinations of $f_\pi$, $f_K$, $f_D$ all agree with experiment.  Figure~\ref{fig:fDs} shows a comparison of recent experimental and theoretical determinations of $f_D$ and $f_{D_s}$. 
This disagreement is somewhat challenging to accommodate in new physics models because Standard Model $D_s$ leptonic decay occurs at tree-level.  Dobrescu and Kronfeld have pointed out, however, that models with a charged Higgs or leptoquark can work~\cite{Dobrescu:2008er}, and that these could also lead to a signal in $D \to K \ell \nu$ semileptonic decay~\cite{Kronfeld:2008gu}.  The only other $N_f = 2+1$ flavor calculation of $f_{D_s}$ by the Fermilab Lattice and MILC collaborations, which was updated in these proceedings~\cite{Simone}, agrees with the CLEO measurement, but has an error that is approximately three times larger than that of HPQCD.  The ETM collaboration recently published an $N_f = 2$ determination of $f_{D_s}$ using twisted-mass fermions that agrees with the CLEO measurement~\cite{Blossier:2009bx}, but the result also has larger errors than those of HPQCD.  Alternative  lattice calculations of $f_D$ and $f_{D_s}$ with comparable precision to that of HPQCD are therefore necessary to either confirm or refute the result.  On the experimental side, BES-III should measure the $D$- and $D_s$-meson leptonic decay constants with $\sim$1\% precision after 4 years of running at their nominal luminosity~\cite{Li:2008wv}, and will help shed light on this $2.4\sigma$ tension.

It is worth noting that the HPQCD collaboration recently obtained a new value for the absolute scale $r_1$ which they use to convert lattice quantities into physical units.  Their published result for $f_{D_s}$ relies upon the older value of $r_1 = 0.321(5)$ fm from the $\Upsilon$ 1S-2S mass-splitting~\cite{Gray:2005ur}.  Their new $r_1 = 0.3133(23)(3)$ fm is $\sim 1.5\sigma$ lower, and comes from averaging the value of $r_1$ obtained from three different methods~\cite{McNeile:2009eq}.  Because the uncertainty in $f_{D_s}$ is currently dominated by the uncertainty in $r_1$, use of the new scale determination will likely increase the value of $f_{D_s}$ by approximately 1.5$\sigma$, or 5 MeV, and thereby reduce the tension with the CLEO measurement to below 2.0$\sigma$.  Thus the ``$f_{D_s}$ puzzle'' may not, in fact, be a puzzle for much longer.

\subsection{$B \to \tau \nu$ leptonic decay}

\begin{figure}[t]
\begin{center}
\includegraphics[height= 0.36\linewidth]{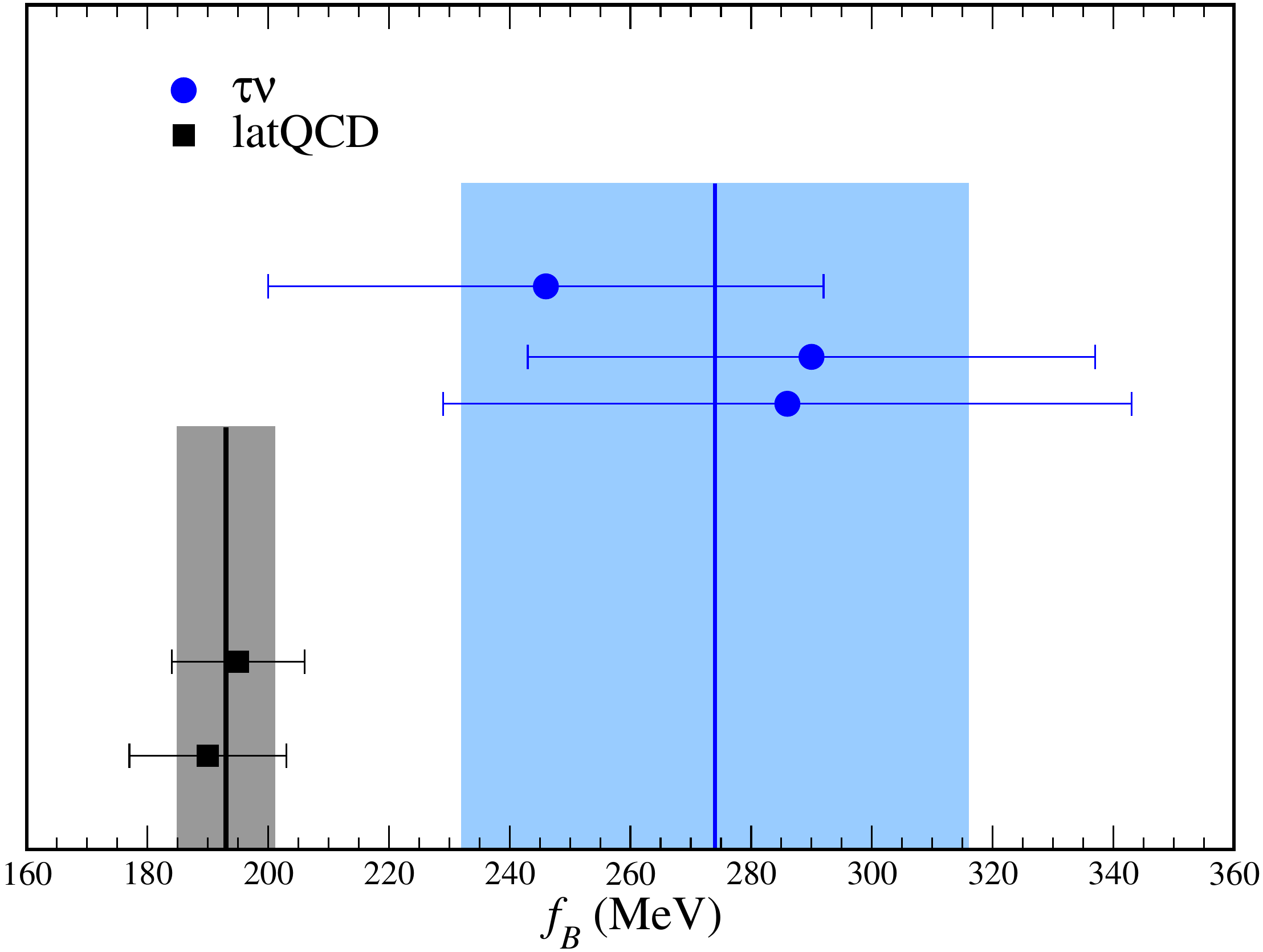}
\includegraphics[height= 0.37\linewidth]{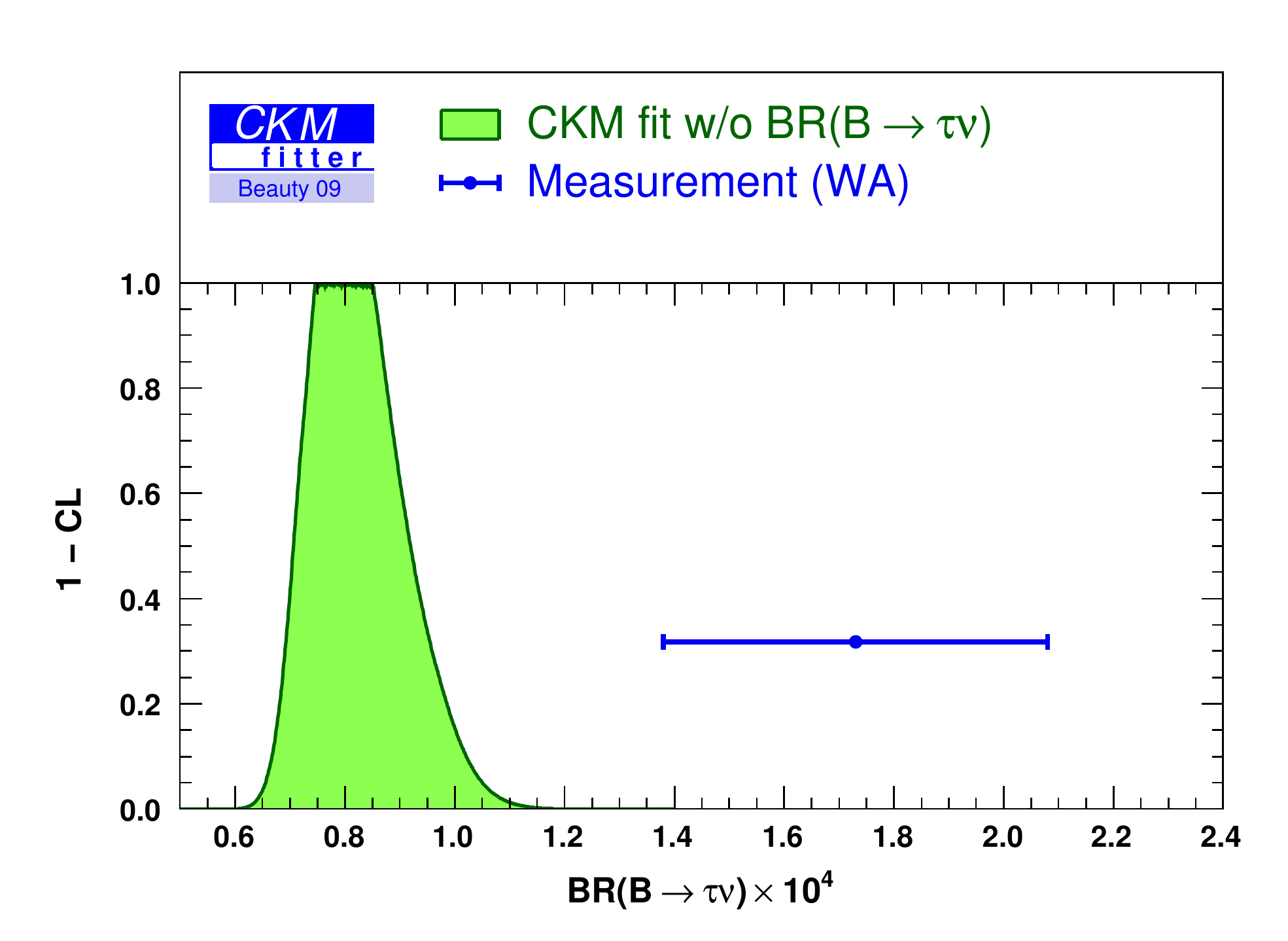}
\caption{Tension between the Standard Model prediction for $B \to \tau \nu$ leptonic decay and current experimental observations.  The left plot compares lattice QCD calculations of $f_B$ with experimental determinations of $f_B$ (using $|V_{ub}|$ from lattice QCD)~\cite{Kronfeld_Pheno09}; the discrepancy is $1.9 \sigma$.  The right plot compares the Standard Model prediction for the $B \to \tau \nu$ branching fraction obtained from the global unitarity triangle fit with the experimentally-measured world average~\cite{CKMfitter_Btotaunu}; the discrepancy is $2.4 \sigma$.}
\label{fig:BtoTaunu}
\end{center}
\end{figure}

There is a $1.9 \sigma$ difference between lattice calculations of $f_B$ and experimental determinations of $f_B$ (using $|V_{ub}|$ exclusive from lattice QCD), as shown in the left-hand plot in Fig.~\ref{fig:BtoTaunu}~\cite{Kronfeld_Pheno09}.
This discrepancy is unrelated to the $f_{D_s}$ puzzle discussed in the previous subsection.  The HPQCD collaboration's calculation of $f_B$~\cite{Gray:2005ad} uses NRQCD $b$-quarks, and is independent of their determination of $f_D$ and $f_{D_s}$ using HISQ charm quarks.  Furthermore, this difference is also observed by the Fermilab Lattice and MILC collaborations, whose determination of $f_B$ has a smaller error than that of HPQCD~\cite{Bernard:2009wr}.  It should be noted both the HPQCD and Fermilab/MILC determinations of $f_B$ will increase slightly when they are recomputed using updated values of the scale $r_1$~\cite{McNeile:2009eq,Bernard:2007ps}.  This should not, however, affect the tension significantly because the uncertainty in $f_B$ from $r_1$ is much less than the total error.

There is also a $2.4$--$2.5\sigma$ discrepancy between the Standard Model prediction for the $B \to \tau \nu$ branching fraction from the global unitarity triangle fit and the experimentally-measured world average~\cite{CKMfitter_Btotaunu,Bona:2009cj},  shown in the right-hand plot in Fig.~\ref{fig:BtoTaunu}.  This comparison relies on several lattice QCD inputs to the various unitarity triangle constraints.  It can be made independent of the theoretical uncertainty in $f_B$, however, by using the ratio of ${\mathcal{B}}(B \to \tau \nu)/\Delta m_d$.  The CKMfitter group uses this ratio to make a prediction for the $B_d$-meson mixing parameter $\hat{B}_{B_d}$.  They obtain $\hat{B}_{B_d} = 0.51^{+0.16}_{-0.10}$~\cite{CKMfitter_FPCP09_Btotaunu}, which differs considerably from the only $N_f=2+1$ flavor lattice result $\hat{B}_{B_d} = 1.26 \pm 0.11$~\cite{Gamiz:2009ku}.

The discrepancy in $B \to \tau \nu$ leptonic decay could be due to the presence of a charged Higgs, but this is not a particularly natural explanation.  Although an enhancement over the Standard Model branching fraction could occur for very light values of the charged Higgs mass, most MSSM scenarios would lead to a suppression of the branching fraction~\cite{Kamenik:2008tj,Antonelli:2009ws}.  Reducing the experimental errors in ${\mathcal{B}}(B \to \tau \nu)$ will be difficult at LHCb, but prospects are good for a super-$B$ factory~\cite{Browder:2008em}.  The theoretical errors in $f_B$ will improve more quickly.  The Fermilab Lattice and MILC collaborations are reducing their errors by adding statistics, while the HPQCD collaboration is exploring the use of HISQ light quarks in combination with NRQCD $b$-quarks~\cite{Shigemitsu:2009jy}.  Ongoing work by the RBC and UKQCD collaborations using domain-wall light quarks and static $b$-quarks~\cite{Witzel} and by the ETM collaboration using twisted-mass light quarks and static $b$-quarks~\cite{Lubicz_poster} was also presented at this conference.

\section{Lattice flavor physics beyond the CKM matrix}
\label{sec:LQCD_rare}

Because current observations are consistent with the Standard Model CKM framework with only a few 2--3$\sigma$ deviations that may or may not turn out to be significant when experimental and theoretical errors improve, it is clear that we should look for new physics in other realms of the quark flavor sector beyond the CKM matrix.  One set of processes that are particularly sensitive to physics beyond the Standard Model are ``rare'' decays,  which are rare because they are suppressed at lowest-order in the the Standard Model.  This can be because the decay is loop-suppressed (as in flavor-changing neutral currents), the decay is helicity-suppressed by powers of small masses, or the decay is ``Cabibbo''-suppressed by powers of small CKM matrix elements.  

The potential to search for new physics in rare decays is far from exhausted.  The Super-$B$ factory, ``Project X'' at Fermilab, and other future experiments will improve measurements of (or observe for the first time) many rare $B$- and $K$-decays.  In this section I therefore discuss ways in which the lattice community can aid in the search for new physics in rare decays.  

\subsection{Rare $b \to d$ and $b \to s$ transitions}

\begin{figure}[t]
\begin{center}
\includegraphics[width= 0.7\linewidth]{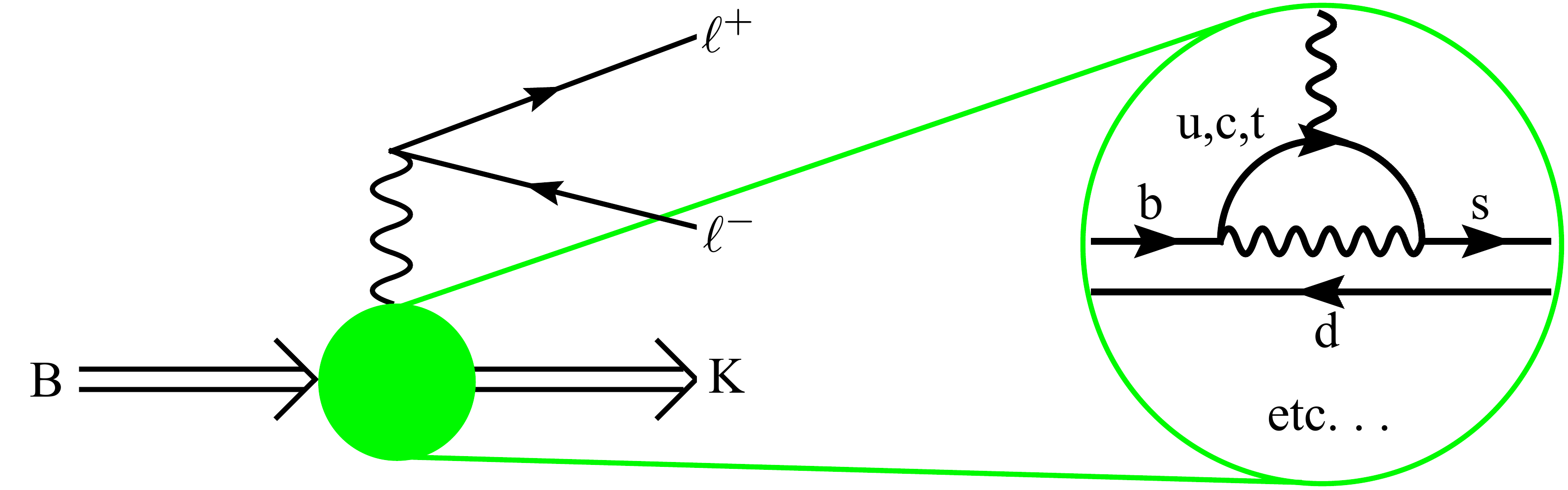}
\caption{Standard Model contribution to $B \to K \ell \ell$ semileptonic decay.  This process is suppressed in the Standard Model because it proceeds through flavor-changing neutral currents.}
\label{fig:btos}
\end{center}
\end{figure}

\begin{figure}[t]
\begin{center}
\includegraphics[width= 0.6\linewidth]{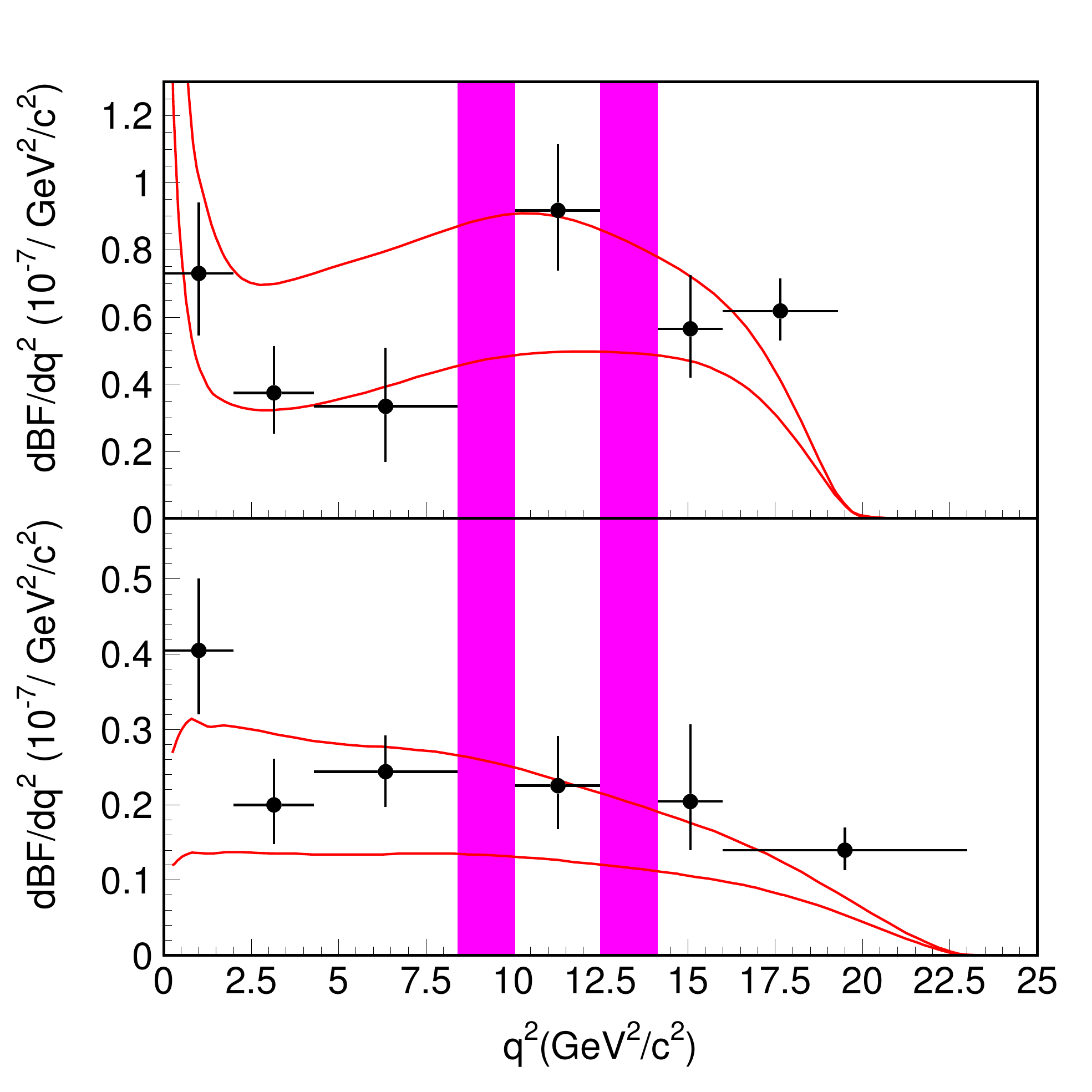}
\caption{Differential branching fractions for $B \to K^* \ell^+ \ell^-$ (upper plot) and $B \to K \ell^+ \ell^-$ (lower plot) versus $q^2$~\cite{:2009zv}. The two shaded regions are veto windows to reject $J/\psi$ and $\psi(2S)$ events. The solid curves show the Standard Model predictions.  The widths of the allowed ranges are dominated by the theoretical uncertainties in the hadronic form factors~\cite{Ali:1999mm,Ali:2002jg}.}
\label{fig:BtoKll}
\end{center}
\end{figure}

Semileptonic decays involving $b \to d$ and $b \to s$ transitions are rare because the lowest-order Standard Model contributions are 1-loop and hence suppressed.  For example, Fig.~\ref{fig:btos} shows a sample Standard Model contribution to $B \to K \ell \ell$ decay.  
Because new particles will enter $b \to d$ and $b \to s$ electroweak penguin loops in many models, new physics contributions to these processes could be of the same size or larger than those from the Standard Model.  Thus $b \to d$ and $b \to s$ transitions are potentially stronger probes of new physics than $b \to u$ transitions.  Figure~\ref{fig:BtoKll} shows the differential branching fractions for  $B \to K^* \ell^+ \ell^-$ and $B \to K \ell^+ \ell^-$ versus $q^2$ measured by the Belle collaboration~\cite{:2009zv}.
Although the errors are still rather large, both LHCb and Super-$B$ will soon improve the measurement of these and other rare $B$-decays. 

The Standard Model predictions for the $B \to K^{(*)} \ell \ell$ branching fractions (denoted by the red solid curves in Fig.~\ref{fig:BtoKll}) are limited by the uncertainty in the hadronic form factors.  Estimates from light-cone sum rules typically lead to $\sim$30\% hadronic uncertainties in the branching ratios~\cite{Buchalla:2000sk}.  For example, in the case of $B \to K \ell \ell$, the Standard Model prediction is ~\cite{Ali:2002jg}
\begin{equation}
{\mathcal{B}}(B \to K \ell^+ \ell^-) = (0.35 \pm 0.11 \pm 0.04 \pm 0.02 \pm 0.0005) \times 10^{-6} \,,
\end{equation}
where the first error is the contribution from the hadronic form factors.
Because the branching fraction is proportional to the square of the hadronic form factors, a reduction in the errors on $f_+(q^2)$ and $f_T(q^2)$ to below 10\% would reduce the total theoretical error significantly.  Lattice QCD therefore has a good opportunity to improve the theory of rare $B$-decays in the Standard Model.  Once the hadronic matrix elements are computed precisely, they can also be combined with the perturbatively-calculated Wilson coefficients to make predictions for beyond-the-Standard Model scenarios and help distinguish between new physics models.

Currently there are no published $N_f = 2+1$ flavor lattice QCD calculations of semileptonic form factors for rare $B$-decays, although work is in progress using staggered light quarks and moving NRQCD heavy quarks~\cite{Meinel:2008th,Liu:2009dj}.  As Fig.~\ref{fig:BtoKll} shows, in the case of $B \to K^{(*)} \ell \ell$ the full $q^2$ range is experimentally accessible (with the exception of the $J/\psi$ and $\psi(2S)$ peaks).  Thus the fact that standard lattice QCD methods can only obtain semileptonic form factors at high $q^2$ is less problematic than in the case of $B\to\pi\ell\nu$ decay.   Furthermore, model-independent formulas based on analyticity and crossing symmetry such as the ``$z$-parameterization'' can be used to extrapolate lattice form factor results to low-$q^2$ in a clean way~\cite{Arnesen:2005ez,Becher:2005bg,Bourrely:2008za}.  This method has already been employed in the case of $B \to \pi \ell \nu$ by the Fermilab Lattice and MILC collaborations~\cite{Bailey:2008wp}.  A challenge for lattice calculations of rare $B$-decays is the fact that many interesting processes such as $B \to K^* \ell \ell$ and $B \to K^* \gamma$ involve vector mesons.  In these cases chiral perturbation theory cannot be used to guide the chiral extrapolation.  Fortunately, this situation will improve as the simulated quark masses continue to become lighter.

\subsection{$K \to \pi \nu \bar\nu$ decays}

\begin{figure}[t]
\begin{center}
\includegraphics[width= 0.7\linewidth]{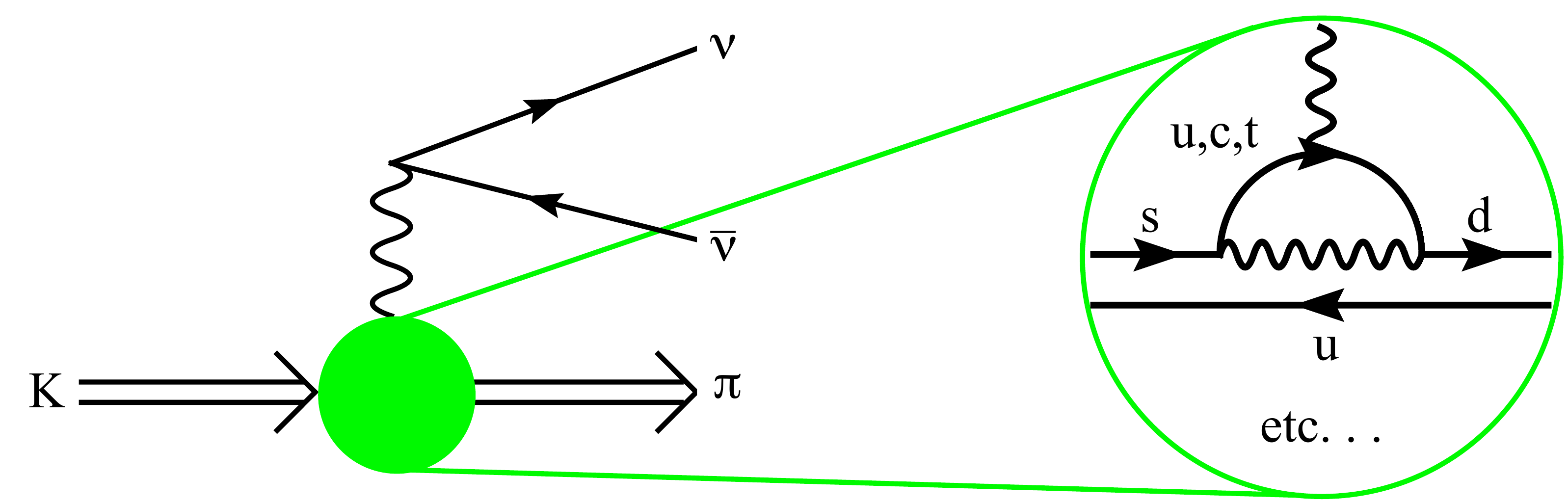}
\caption{Standard Model contribution to $K \to \pi \nu \bar\nu $ semileptonic decay.  This process is suppressed in the Standard Model because it proceeds through flavor-changing neutral currents.}
\label{fig:Ktopinunubar}
\end{center}
\end{figure}

The semileptonic decay $K \to \pi \nu \bar\nu$ proceeds through flavor-changing neutral currents, as shown in Fig.~\ref{fig:Ktopinunubar}, so its lowest-order contributions are 1-loop and hence small in the Standard Model.
In fact, $K \to \pi \nu \bar\nu$ decays are so rare that, although the process has been observed in the $K^+$ channel~\cite{Artamonov:2008qb}, only a limit has been set in the more challenging $K^0$ channel~\cite{Ahn:2007cd}:
\begin{eqnarray}
\mathcal{B}(K^+ \to \pi^+ \nu \bar\nu)_\textrm{exp.} &=& 1.73^{+1.15}_{-1.05} \times 10^{-10} \,,  \\
\mathcal{B}(K^0_L \to \pi^0 \nu \bar\nu)_\textrm{exp.} &<& 670 \times 10^{-10} \quad (90\% \textrm{CL}) \,.
\end{eqnarray}
Both of these results will improve, however, in the near future.  Experiment NA62 at the CERN SPS will measure $\CO$(100) events in the $K^+$ channel, while E391a at KEK (to become E14 at J-PARC) will collect the first $K^0$ events.  Thus it is important to consider how lattice QCD calculations can aid in the search for new physics in rare $K$-decays.  

\begin{figure}[t]
\begin{center}
\includegraphics[height= 0.4\linewidth]{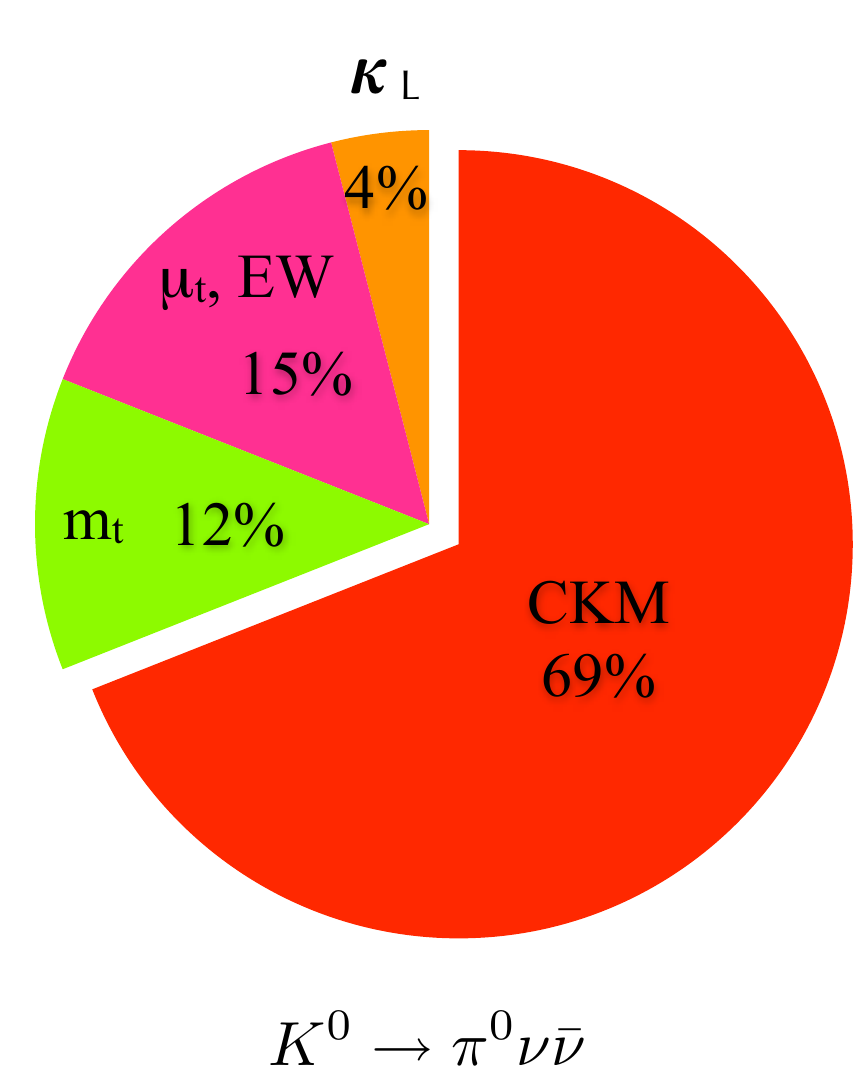} \qquad \qquad \qquad
\includegraphics[height= 0.4\linewidth]{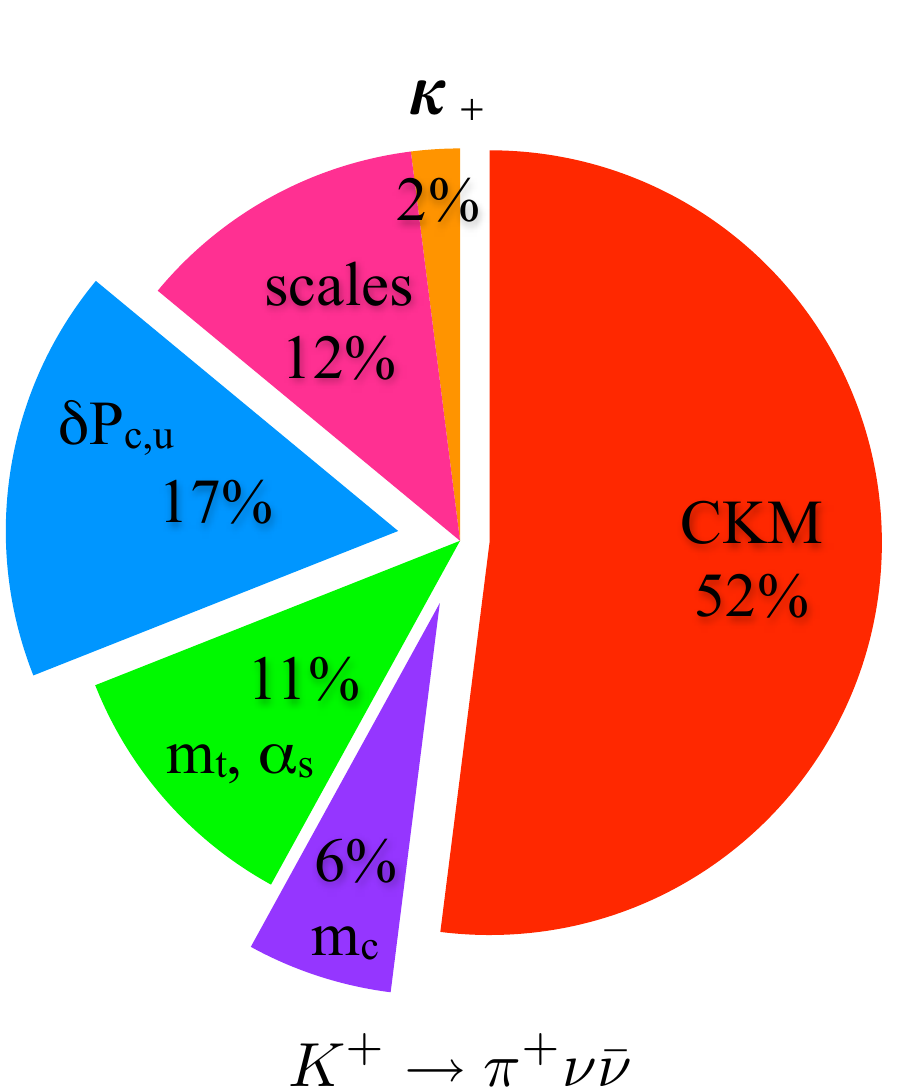}
\caption{Error budgets for the Standard Model predictions of ${\mathcal{B}}(K_L^0 \to \pi^0 \nu \bar\nu) $ (left plot) and ${\mathcal{B}}(K^+ \to \pi^+ \nu \bar\nu)$ (right plot).  Details are described in the text.  Figures courtesy of U. Haisch~\cite{Haisch:2007pd}.}
\label{fig:Ktopinunu_pies}
\end{center}
\end{figure}

The Standard Model branching fractions for $K \to \pi \nu \bar\nu$ are theoretically under much better control than in the case of rare $B$-decays, and are known to 10-15\% accuracy~\cite{Mescia:2007kn,Brod:2008ss}:
\begin{eqnarray}
{\mathcal{B}}(K^0_L \to \pi^0 \nu \bar\nu)_\textrm{SM} &=& (0.249 \pm 0.039) \times 10^{-10} \,, \\
{\mathcal{B}}(K^+ \to \pi^+ \nu \bar\nu)_\textrm{SM} &=& (0.85 \pm 0.07) \times 10^{-10}  \,.
\end{eqnarray}
Furthermore, the uncertainties in the hadronic form factors lead to only a tiny percentage of the total error.  Figure~\ref{fig:Ktopinunu_pies} shows the theoretical error budgets for the Standard Model branching fractions in the $K^0$ and $K^+$ channels;  the form factor $\kappa_L$ contributes 4\% of the error in ${\mathcal{B}}(K_L^0 \to \pi^0 \nu \bar\nu)$, while the form factor $\kappa_+$ contributes 2\% of the error in ${\mathcal{B}}(K^+ \to \pi^+ \nu \bar\nu)$.
This is because the form factors can be estimated quite precisely using experimental $K \to \pi \ell \nu$ data combined with NLO and partially-NNLO $\chi$PT to account for isospin-breaking effects~\cite{Mescia:2007kn}.  

Despite the fact that the $K \to \pi \nu \bar\nu$ hadronic form factors are already well-known, lattice QCD can still help reduce the theoretical uncertainties in the Standard Model branching fractions in several ways.  The dominant uncertainties in the $K^0$ and $K^+$ branching fractions are parametric errors from the CKM matrix elements.   Because the branching fractions are proportional to the Wolfenstein parameter $A^4$, lattice QCD can make a significant impact by reducing the uncertainty $|V_{cb}|$.  The subleading errors in ${\mathcal{B}}(K_L^0 \to \pi^0 \nu \bar\nu)$, however, are from the uncertainty in the top quark mass and unknown higher-order corrections to the Inami-Lim function $X(x_t)$, and must be addressed with better measurements of $m_t$ and higher-order perturbative calculations of $X(x_t)$.  The second-largest error in ${\mathcal{B}}(K^+ \to \pi^+ \nu \bar\nu)$ is due to the long-distance contribution from up and charm quarks.  Isidori, Martinelli, and Turchetti have proposed a method for computing $\delta P_{c,u}$ using lattice QCD that can reduce the contribution to the error on the rate to the 1-2\% level~\cite{Isidori:2005tv}.  The error budget for ${\mathcal{B}}(K^+ \to \pi^+ \nu \bar\nu)$ presented in Fig.~\ref{fig:Ktopinunu_pies} uses the charm quark mass recently obtained from lattice QCD calculations of current-current correlators by the HPQCD collaboration~\cite{Allison:2008xk};  thus the 6\% error can be reduced even further with improved lattice determinations of $m_c$.  The remaining errors in ${\mathcal{B}}(K^+ \to \pi^+ \nu \bar\nu)$, however, can only be addressed with better measurements of $m_t$, more precise determinations of $\alpha_s$, and higher-order perturbative calculations of $X(x_t)$.

\begin{figure}[t]
\begin{center}
\includegraphics[width= 1.0\linewidth]{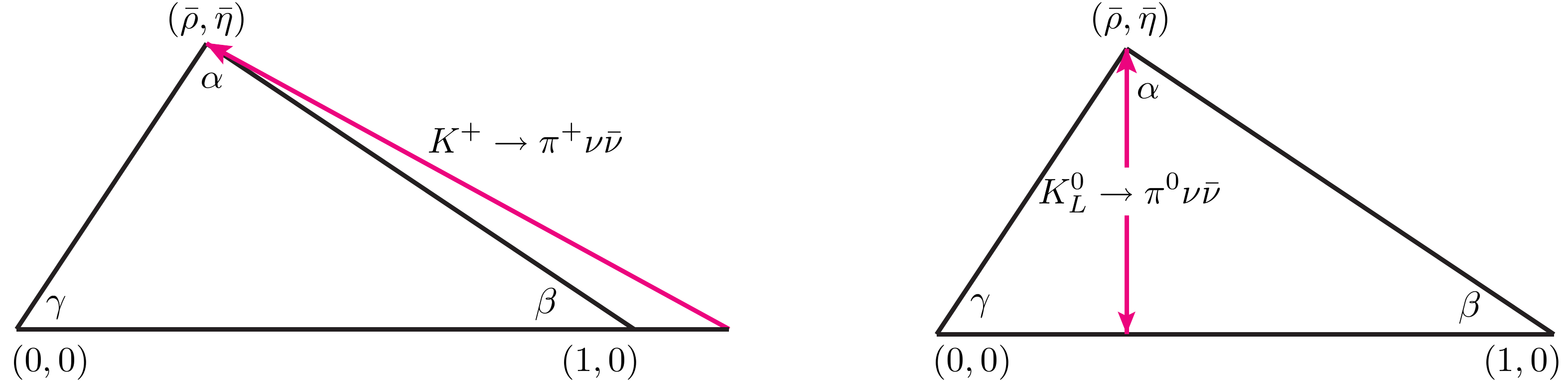}
\caption{Constraints on the CKM unitarity triangle from $K^+\to \pi^+ \nu\bar\nu$ (left plot) and $K^0_L \to \pi^0 \nu \bar\nu$ (right plot).}
\label{fig:Ktopinunu_UT}
\end{center}
\end{figure}

Once the $K \to \pi \nu \bar\nu$ branching fractions have been measured more precisely, they can be used to determine the apex of the CKM unitarity triangle:
\begin{eqnarray}
{\mathcal{B}}(K^+ \to \pi^+ \nu \bar\nu) &=& \kappa_+ A^4 X(x_t)^2 \frac{1}{1 + \lambda^2} \times \left[  (1 + \lambda^2)^2 \bar\eta^2 + (1 + \frac{P_0}{A^2 X(x_t)} - \bar\rho)^2 \right] \,, \\
{\mathcal{B}}(K_L^0 \to \pi^0 \nu \bar\nu) &=& \kappa_L A^4 \bar\eta^2 X(x_t)^2 \,.
\end{eqnarray}
These constraints are shown in Fig.~\ref{fig:Ktopinunu_UT}.
The location of the apex of the CKM unitarity triangle as determined strictly from kaons can then be compared with one from clean $B$-physics observables like $\sin(2\beta)$ and neutral $B$-mixing.  This will provide a highly non-trivial test of the CKM picture.  Even if the $K$- and $B$-physics predictions agree, rare kaon and $B$-decays still probe new physics operators at effective scales up to several TeV or higher.  Thus they will play an important role in discriminating between Standard Model extensions.  

\section{Summary and outlook}
\label{sec:Conc}

Lattice QCD and flavor physics are entering a mature era in which experimental measurements and theoretical calculations are sufficiently precise to constrain the presence of new physics in the flavor sector.  In particular, lattice QCD can now be used reliably compute hadronic matrix elements that encode the nonperturbative QCD contributions to weak processes.  For many of the weak matrix elements of interest, there are now or soon will be several independent $N_f = 2$ or $N_f = 2+1$ flavor lattice calculations, thereby lending credibility to the results~\cite{Aubin:2009yh,Lubicz_plenary,Scholz:2009yz,Laiho:2009eu}.  Although significant progress in reducing the lattice errors is being made, many of the CKM matrix element determinations and most of the constraints on the CKM unitarity triangle are still limited by lattice uncertainties, and it will take several more years to obtain percent-level accuracy in many quantities.

Current observations are consistent with Standard Model expectations that the CKM matrix is unitary to our present level of experimental and theoretical precision.  Although we have not yet found a ``smoking gun'' of new physics in the flavor sector, there are several 2--3$\sigma$ hints that should be monitored, such as the tension between the $\varepsilon_K$ band and the remaining CKM unitarity triangle constraints and the $f_{D_s}$ puzzle.  New physics may not appear first, however, in the gold-plated processes listed in Fig.~\ref{tab:CKM_decays} that are the focus of most lattice efforts.  Thus, given the mature status of many lattice weak matrix element calculations, it is now time to expand the standard lattice QCD repertoire to include rare $K$- and $B$-decays, beyond-the-Standard Model contributions to neutral kaon and $B$-mixing,  and other even more challenging quantities such as $K \to \pi \pi$ decay, $D^0$-$\bar{D^0}$ mixing, and the neutron electric dipole moment.  The lattice community must be prepared for new physics wherever it may arise in the flavor sector so that we can take advantage of this exciting opportunity to find physics beyond the Standard Model.  

\section*{Acknowledgments}

I would like to thank G. Buchalla, A. El-Khadra, U. Haisch, A. Kronfeld, J. Laiho, E. Lunghi, P. Mackenzie, S. Sharpe, B. Tschirhart, and M.~Wingate for help in the preparation of the talk and proceedings.  I would also like to thank  A. Kronfeld and S. Sharpe for providing useful comments on the manuscript.  This manuscript has been authored by employees of Brookhaven Science Associates, LLC under Contract No. DE-AC02-98CH10886 with the U.S. Department of Energy.


\end{document}